\documentstyle[preprint,eclepsf,seceq]{jpsj} 
\def\runtitle{Induced current around the non-magnetic impurity}
\def\runauthor{Yukihiro {\sc Okuno}} 

\title
{Current patterns and magnetic impurities 
 in time-reversal breaking 
superconductors }

\author
{Yukihiro {\sc Okuno}\footnote{E-mail: okuno@yukawa.kyoto-u.ac.jp}}

\inst
{ Yukawa Institute for Theoretical Physics, Kyoto University, Kyoto 
606-8502, Japan }

\abst
{ 
We study the impurity effect in the time reversal 
symmetry (${\cal T}$) breaking superconductor based on the 
Bogoliubov-de Gennes (BdG) equations.
In ${\cal T}$-violating superconductors, 
spontaneous currents are  induced around the impurity. 
The current patterns around the impurity reflect the 
structures of the Cooper pairs. We investigate 
impurity problem numerically for  
two  kinds of  ${\cal T}$ violating superconductors 
$(p_{x}+{\rm i}p_{y}$ and $d+{\rm i}s)$ 
and 
investigate the currents around the impurity. We also 
study the effects of the magnetic impurity in p-wave 
($p_{x}+{\rm i}p_{y}$) superconductor,
especially 
in view of the zero-energy crossing of energy levels 
related to the phase transition of the ground state.
}

\kword
{
time-reversal breaking state, $p_{x}+{\rm i}p_{y}$ and 
$d+{\rm i}s$ superconductor, 
B-dG equation,
spontaneous current, bound state, magnetic impurity
}
\begin{document}
\sloppy
\maketitle
\section{Introduction}
One of the current important issues of the superconductivity is 
the occurrence of the time-reversal symmetry
(${\cal T}$)-violating state.\cite{rf:Sigrist}
It has been investigated  in connection with 
the  local behavior of the order parameter
at the interface of the Josephson junctions and the surface 
of the material. 
Another context of ${\cal T}$ violating state is the 
unconventional superconductor where time reversal symmetry 
is violated throughout the whole materials. Such  examples 
are seen in the heavy fermion superconductors  UPt${}_{3}$,  
U${}_{1-x}$Th${}_{x}$Be${}_{13}$
$(0.017 < x < 0.045)$~\cite{rf:Heffner} and the 
layered perovskite superconductor 
Sr${}_{2}$RuO${}_{4}$~\cite{rf:Maeno}.
In general the violation of time-reversal invariance is 
a property of the magnetism and indeed, it is found that 
the ${\cal T}$-violating superconductor shows various magnetic 
properties.~\cite{rf:Volovik,rf:Sigrist2}
In such systems, inhomogeneities  where the Meissner screening 
effects are insufficient allow the appeared 
 spontaneous supercurrent.  
This happens, for example, at the surface of samples and also 
on the domain wall between the degenerate states of the 
superconductor. The defects of the crystal lattice, 
in particular impurities, can also lead to the appearance of 
the magnetic properties of the ${\cal T}$-violating state, and 
it is known that the spontaneous suppercurrents are 
induced in the vicinity of an impurity in such a systems.
Such an induced current around the impurities can generate the 
local magnetic field in the superconducting state. 
The local magnetic field can be detected by 
a high resolution probe 
like spin polarized muons~\cite{rf:Heffner,rf:Luke}
 and 
indeed has lead to the observation of the intrinsic 
magnetism in the superconducting state 
of such ${\cal T}$-violating superconducting systems.
There are some studies about the induced current 
around the impurity in ${\cal T}$-violating 
superconductor~\cite{rf:Choi,rf:Mineev,rf:Rainer,rf:Okuno}.
We have already investigated a microscopic mechanism of 
such an induced current around the 
non-magnetic impurity with the Cooper pair of the type 
$p_{x}\pm {\rm i}p_{y}$, which is a candidate of 
the Cooper pair symmetry of the 
Sr${}_{2}$RuO${}_{4}$~\cite{rf:Okuno}
 and has angular momentum $L_{z}=\pm 1$.
We see that the circular current 
generated around the impurities 
reflects the angular momentum of the Cooper pair. 
But still there are only few 
studies on how the induced currents around 
the impurity depend on the type of the Cooper pairs. 
So it is interesting to study the influence of the 
the structures of the Cooper pairs to the patterns of the 
induced currents around the impurity.
Here in this paper we investigate the induced currents around 
a non-magnetic impurity with different types of Cooper pair, 
especially we pay attention to the angular momentum of the 
Cooper pair. Basically we can make a distinction between 
two kinds of ${\cal T}$-violating superconducting states 
whose Cooper pair have an intrinsic angular momentum  
like $p_{x}\pm {\rm i}p_{y}$, $d_{zx}\pm {\rm i}d_{zy}$
($L_{z}=\pm 1$), $d_{x^{2}-y^{2}}\pm {\rm i}d_{xy}$ 
($L_{z}=\pm 2$) and those that have no angular momentum like 
$d+{\rm i}s$. Such a difference of the structure of the 
Cooper pairs will affect the pattern of the induced currents 
around the impurity. We also study the effects of the magnetic 
impurity in $p_{x}+{\rm i}p_{y}$ superconductor and 
investigate the differences from the non-magnetic impurity.
Especially we pay attention to the bound state and 
the ground state transition as well as  the induced current. \\
The organization of this paper is as follow. In $\S 2$, 
we solve the Bogoliubov-de Gennes equation numerically and 
investigate the induced current around the impurity.
In $\S 3$, we study the magnetic impurity effect on the 
$p_{x}+{\rm i}p_{y}$ superconductor, $\S 4$ is the 
conclusion and discussion.
\section{Pattern of the induced current around 
a non-magnetic impurity }
In order to investigate the impurity effect on the superconductor, 
we solve the Bogoliubov de-Gennes equation 
numerically~\cite{rf:Onishi,rf:Kusama}. 
The formalism is given below. 
We start with the B.C.S Hamiltonian on the 2D square lattice 
\begin{eqnarray}
H &=& -t\sum_{<i,j>,\sigma }a^{\dag}{}_{i \sigma}a_{j \sigma}  
+\sum_{<i,j>}(
\Delta_{i,j}a^{\dag}{}_{i \uparrow}a^{\dag}{}_{j \downarrow} 
+{\rm h.c} )
+{\rm U}\sum_{\sigma }a^{\dag}{}_{0 \sigma}a_{0 \sigma}
-\mu \sum_{i,\sigma }
a^{\dag}{}_{i \sigma }a^{\dag}{}_{i \sigma } \label{HAM0}
\end{eqnarray}
where $t$ and ${\rm U}$ is the hopping between the nearest-neighbor 
(n.n) sites and the impurity potential respectively and 
$\mu $ is the chemical potential of the system. We set 
the non-magnetic impurity at the site $0$.
The real space order parameter $\Delta_{i,j}$ is defined as 
\begin{eqnarray}
\Delta_{i,j} &=& 
V_{i,j} <a_{i \uparrow}a_{j \downarrow}> \label{gap}
\end{eqnarray}
where $V_{i,j}$ is the attractive pairing interaction between 
the site $i$ and $j$. 
The Hamiltonian (\ref{HAM0}) can be diagonalized by means of 
the unitary transformation ${\cal U}$ 
\begin{eqnarray}
H &=& \sum_{m=1}^{N_{L}}
\epsilon_{m}\gamma^{\dagger }{}_{m \uparrow }
\gamma_{m \uparrow }+
\sum_{m=1}^{N_{L}}
(-\epsilon_{m})\gamma_{m \downarrow }
\gamma^{\dagger}{}_{m \downarrow } \label{HAM2}
\end{eqnarray}
where $N_{L}$ is the number of the lattice sites and 
$\epsilon_{m} ( >0 )$ is the excitation energy of the  
Bogoliubov quasiparticles. The fermion operators $\gamma $ 
are related to the original electron operator 
$a_{i,\sigma }$ by 
\begin{eqnarray}
\left(
\begin{array}{c}
a_{1 \uparrow} \\ \vdots \\
a_{N_{L} \uparrow} \\ a^{\dagger }{}_{1 \downarrow} \\
\vdots \\ a^{\dagger }_{N_{L} \downarrow} 
\end{array} 
\right) &=&
{\cal U}
\left(
\begin{array}{c}
\gamma_{1 \uparrow} \\ \vdots \\
\gamma_{N_{L} \uparrow} \\ 
\gamma^{\dagger }{}_{1 \downarrow} \\
\vdots \\ \gamma^{\dagger }_{N_{L} \downarrow} 
\end{array}
\right) .
\end{eqnarray}
Then, the self-consistent equation for the gap $\Delta_{i,j} $ 
(\ref{gap}) is given as
\begin{eqnarray}
\Delta_{i,j} &=& -\sum_{m=1}V_{i,j}
[{\cal U}^{\ast}{}_{N_{L}+j,N_{L}+m}
{\cal U}_{i,N_{L}+m} ]. \label{gap2}
\end{eqnarray}
We solve  the equation (\ref{HAM0}) - (\ref{gap2}) iteratively 
until the self consistent conditions are satisfied. 
In the ${\cal T}$-violating superconductor, the spatial variation 
of the gap function induces the supercurrent. The current 
in the $a$ directions at the site $i$ is 
expressed as follow,
\begin{eqnarray}
J_{ia} &=& {\rm i}t\sum_{m=1}
({\cal U}_{i+e_{a}+N,m}{\cal U}^{\ast}{}_{i+N,m}-
{\cal U}^{\ast}{}_{i+e_{a}+N,m}{\cal U}{}_{i+N,m}) \nonumber \\
&+&{\cal U}^{\ast}{}_{i+e_{a},m+N}{\cal U}{}_{i,m+N}
-{\cal U}{}_{i+e_{a},m+N}{\cal U}^{\ast}{}_{i,m+N}).
\end{eqnarray}
\\
First, we solve the problem in the case of 
$\eta_{\pm}=p_{x}\pm {\rm i}p_{y}$, which has an 
angular momentum $L_{z}=\pm1$
if the systems has cylindrical symmetry.
We take the nearest neighbor interaction for 
$V_{i,j}$ 
and shift the chemical potential from  half filling
in order to stabilize  the p-wave pairing.
We assume the state 
$\eta_{+}=p_{x}+{\rm i}p_{y}$ is realized globally in the 
absence of impurity.
In the calculation we take the lattice size 27$\times$27 with 
periodic boundary condition and 
 we take the value $V_{i,j}=3t$, 
$U=100t$, the chemical potential $\mu=-0.7t$. \\
The numerical result satisfies the relation,
\begin{eqnarray}
\Delta_{i,i+{\bf e_{x}}} &=& 
-\Delta_{i+{\bf e_{x}},i}, 
\nonumber \\
\Delta_{i,i+{\bf e_{y}}} &=& 
-\Delta_{i+{\bf e_{y}},i}, 
\end{eqnarray}
where ${\bf e_{x}}$ and ${\bf e_{y}}$ are unit vectors
in the 
x and y direction respectively.
This result shows that 
the  odd parity state is realized and there are no 
even parity components within the numerical accuracy.
Each component of the $p_{x}$ and $p_{y}$ 
of the Cooper pairs at the site $i$ 
are defined as,
\begin{eqnarray}
\Delta_{{\rm p}_{x}}(i) &\equiv & 
\frac{1}{2}(\Delta_{i,i+{\bf e_{x}}}-\Delta_{i,i-{\bf e_{x}}}) \\
\Delta_{{\rm p}_{y}}(i) &\equiv & 
\frac{1}{2}(\Delta_{i,i+{\bf e_{y}}}-\Delta_{i,i-{\bf e_{y}}})
\end{eqnarray}
Due to the spatial variation of the order parameter, 
$\eta_{-}=p_{x}-{\rm i}p_{y}$ component is induced near the 
impurity. We determine the weight of the $\eta_{\pm}$ 
for each site 
as, 
\begin{eqnarray}
\eta_{+}(i)\left(
\begin{array}{c}
1\\
{\rm i}
\end{array}
\right)
+
\eta_{-}(i)\left(
\begin{array}{c}
1\\
-{\rm i}
\end{array}
\right) 
&=& 
\left(
\begin{array}{c}
\Delta_{p_{x}}(i)\\
\Delta_{p_{y}}(i)
\end{array}
\right)
\end{eqnarray}
We depict the spatial dependence of the order parameter 
in Fig.1(a-c).
Fig.(1-a) shows the $\eta_{+}$ component which is dominant 
in the bulk and are suppressed around the impurity.
In Fig.1b-c, 
we can see the admixture component $\eta_{-}$,  
which vanishes in the bulk and  is 
induced around the impurity.
The admixture component $\eta_{-}$  has 
phase winding 2, like 
$\eta_{+}(i)-{\rm e}^{2{\rm i}\theta}\eta_{-}(i)$, 
reflecting the difference of the angular momentum between 
dominant and admixture components of the Cooper pairs. 
The BdG equations have bound state which is localized around the 
impurity. The energy level of the bound state is not zero 
even in the unitary limit, if the system has not 
 the electron-hole symmetry (see Appendix 1).
\begin{figure}[t]
   \epsfile{file=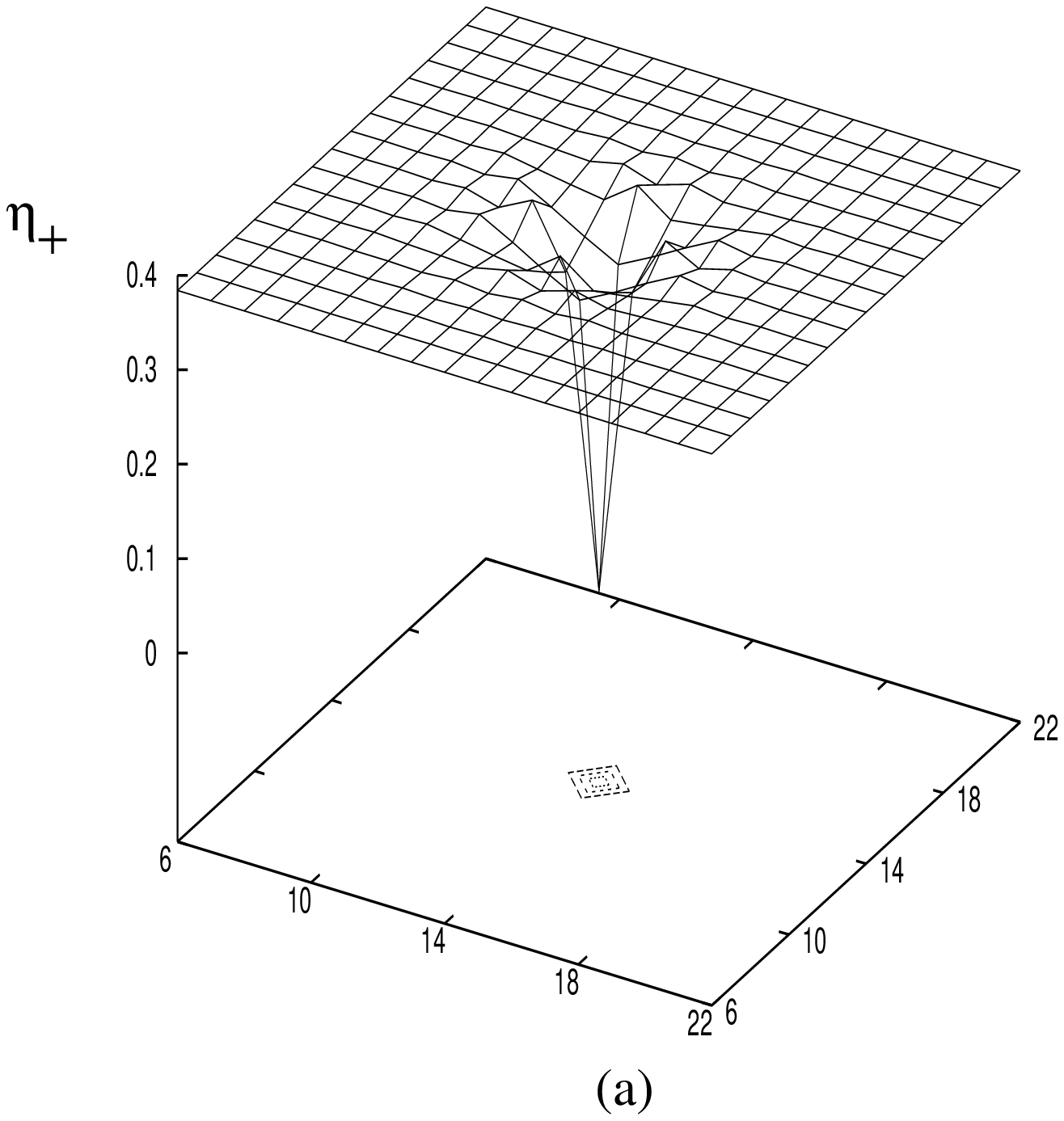,width=6.5cm} 
   \epsfile{file=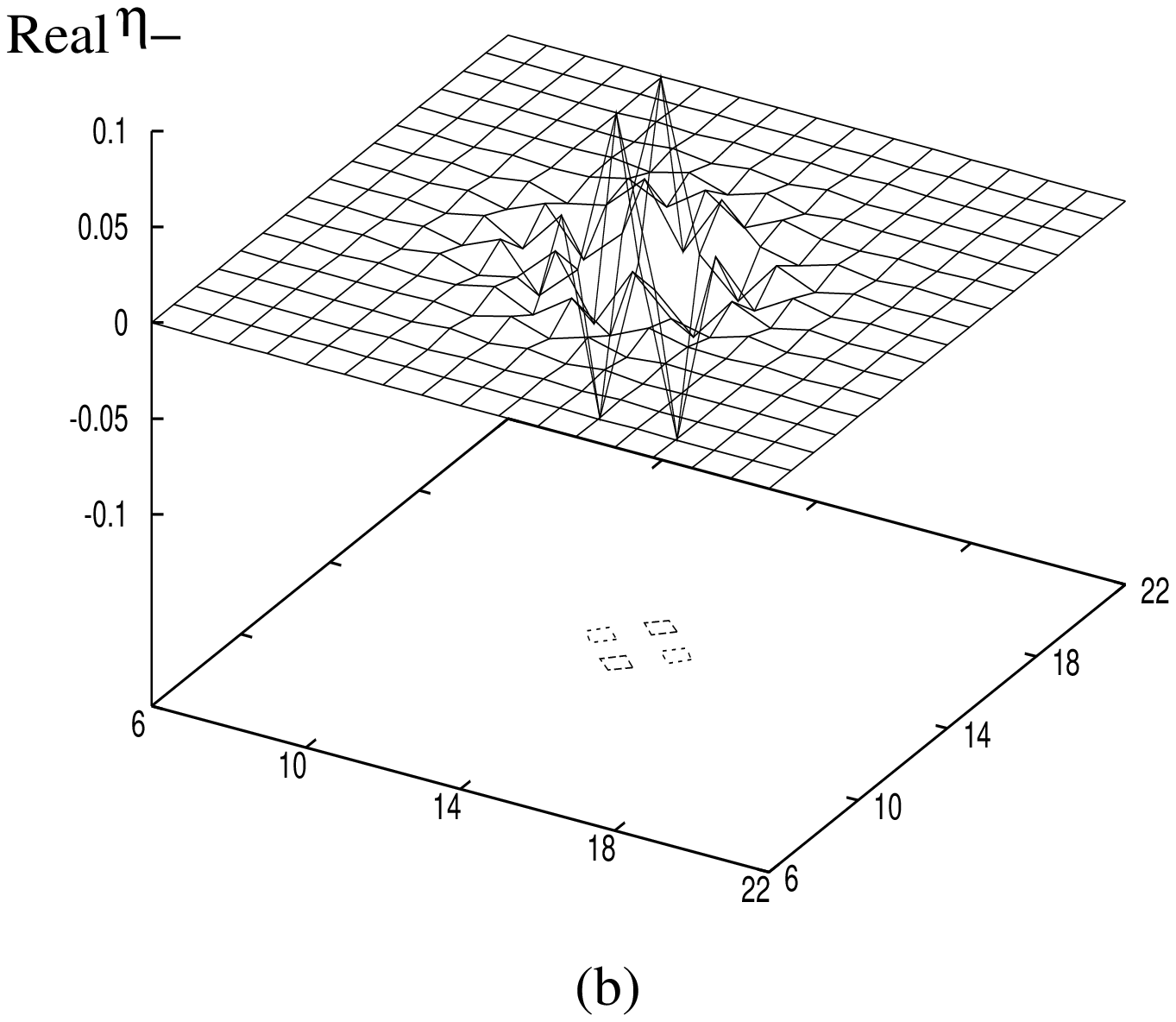,width=6.5cm} \\
   \epsfile{file=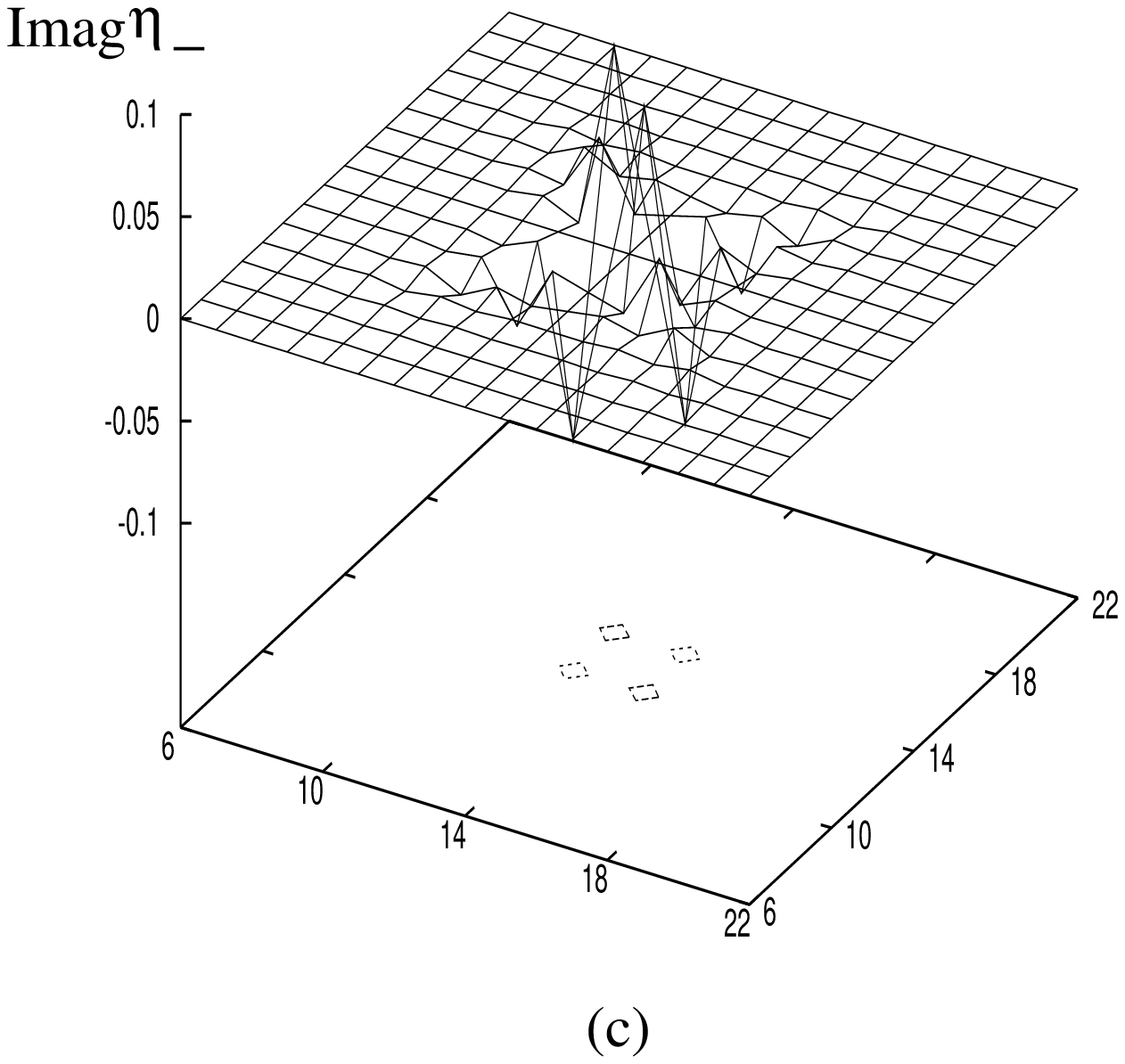,width=6.5cm}
\caption{
The self consistent gap function for the 
dominant  $\eta_{+}(i)$ component and the admixture 
$\eta_{-}(i)$ components. Fig.(1-a) is the spatial variation 
of the $\eta_{+}(i)$ component. Fig.(1-b) and (1-c) are  
 the real and imaginary component of $\eta_{-}(i)$ 
respectively.
}
\label{fig:fig1abc}
\end{figure}

In Fig.2 we show the induced circular current. 
The spontaneous circular current are induced around the impurity 
and the current pattern 
are the reflection of the intrinsic angular momentum of the
Cooper pairs.
The intrinsic angular momentum turns
into a circular current around an impurity due to scattering.
 Within this numerical calculation on the lattice 
systems, it is difficult to observe  the `counter current' 
which we discussed in Ref\cite{rf:Okuno} 
because it is small and extends beyond the lattice treated model. \\
\begin{figure}[t]
\begin{center}
   \epsfile{file=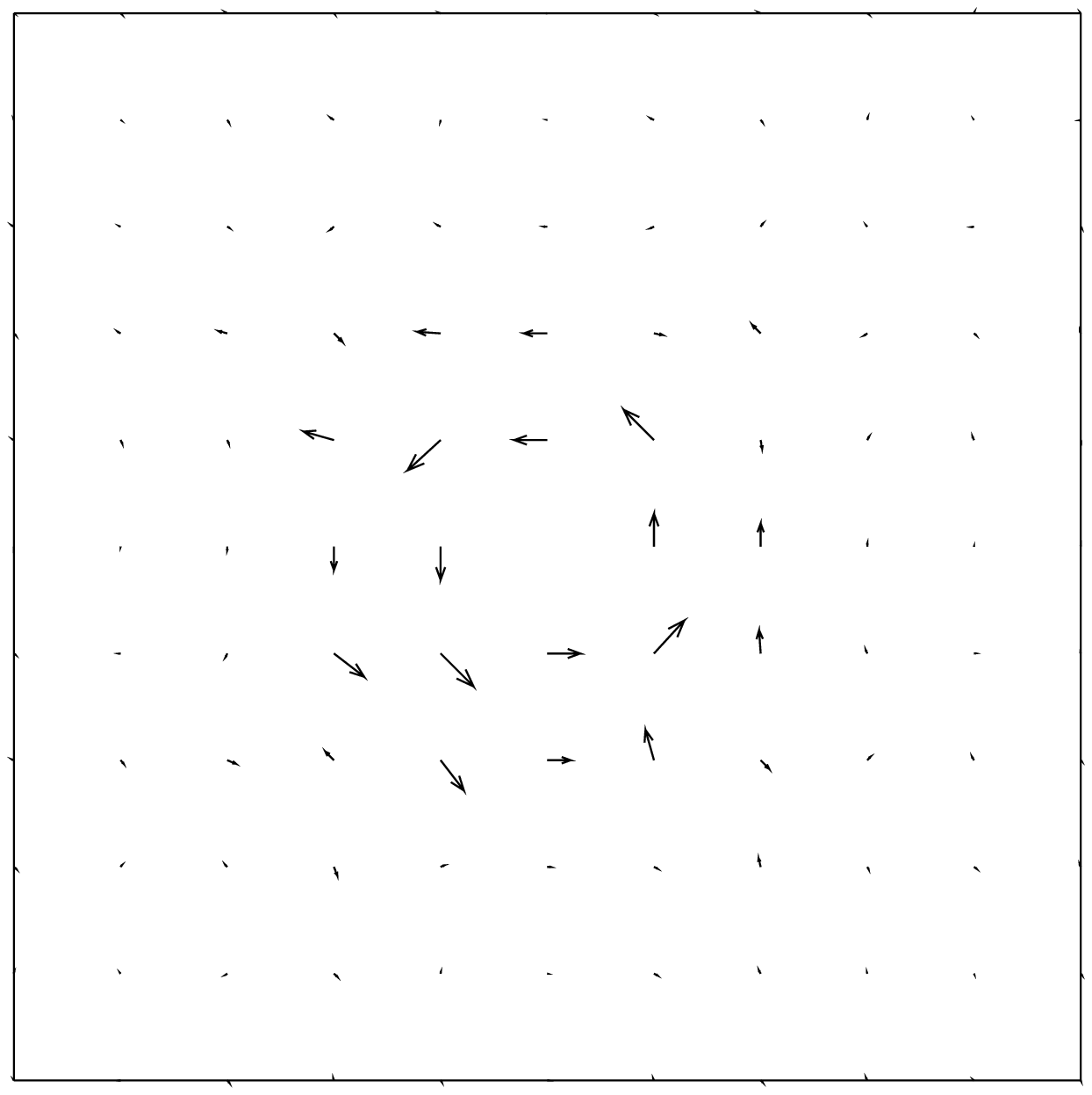,width=5cm}
\end{center} 
\caption{
Induced current around the impurity in
$p_{x}+{\rm i}p_{y}$ superconductor.
}
\label{fig:fig2}
\end{figure}
We also show  another type of the ${\cal T}$-violating 
 superconductor. We take  $d_{xy}+{\rm i}s$, which is 
 easily realized by the next-nearest neighbor interaction. 
 In this case, the Cooper pair has 
 no angular momentum (equal admixture of the orbital 
 momentum $L_{z}=\pm2$) and is not a chiral state.
 We define the d-wave and (extended) s-wave components 
 of the gap function as below,
\begin{eqnarray}
\Delta_{d}(i) &=& 
\frac{1}{4}(\Delta_{i,i+{\bf e_{x}}+{\bf e_{y}}}
+\Delta_{i,i-{\bf e_{x}}-{\bf e_{y}}}
-\Delta_{i,i-{\bf e_{x}}+{\bf e_{y}}}
-\Delta_{i,i+{\bf e_{x}}-{\bf e_{y}}}) \nonumber \\
\Delta_{s}(i) &=& 
\frac{1}{4}(\Delta_{i,i+{\bf e_{x}}+{\bf e_{y}}}
+\Delta_{i,i-{\bf e_{x}}-{\bf e_{y}}}
+\Delta_{i,i-{\bf e_{x}}+{\bf e_{y}}}
+\Delta_{i,i+{\bf e_{x}}-{\bf e_{y}}}) 
\end{eqnarray}
In the calculation we take the next-nearest 
neighbor interaction 
$V_{i,j}=V=2t$ and the chemical potential $\mu=-0.7t$ as 
in the p-wave case.  
In Fig.3(a-c) we show the self-consistently determined 
order parameters. The magnitude of the bulk $d_{xy}$ 
and $s$-wave component is estimated as $\Delta_{d}=0.6t$ and 
$\Delta_{s}=0.22t$ within these parameters. 
Around the impurity 
the real parts of the s-wave order is generated.
Not that also the s-wave component is influenced by the non-magnetic 
impurity, since it has a non-trivial phase dependence in the 
Brillouin zone.
\begin{figure}[t]
   \epsfile{file=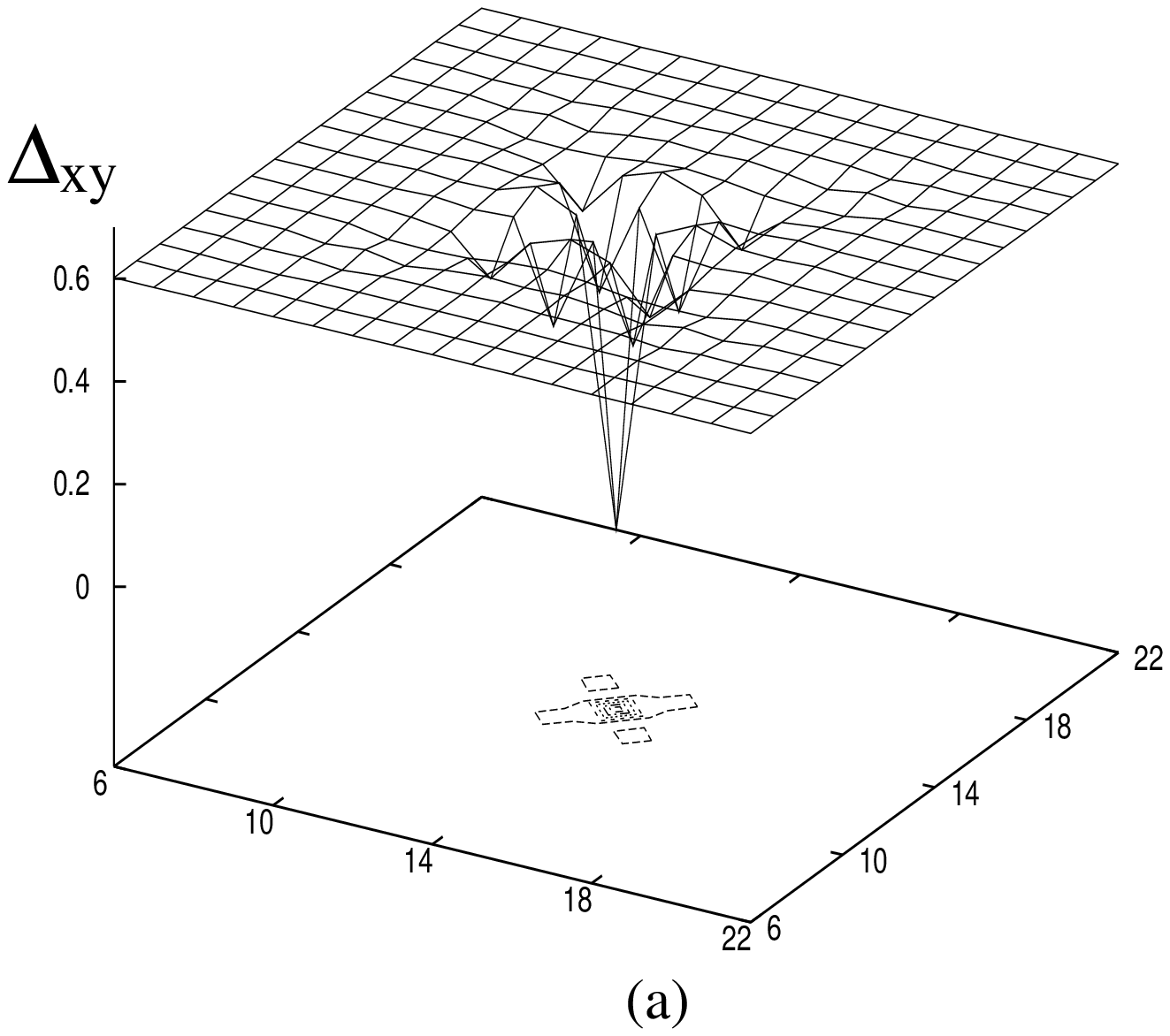,width=6.5cm}
   \epsfile{file=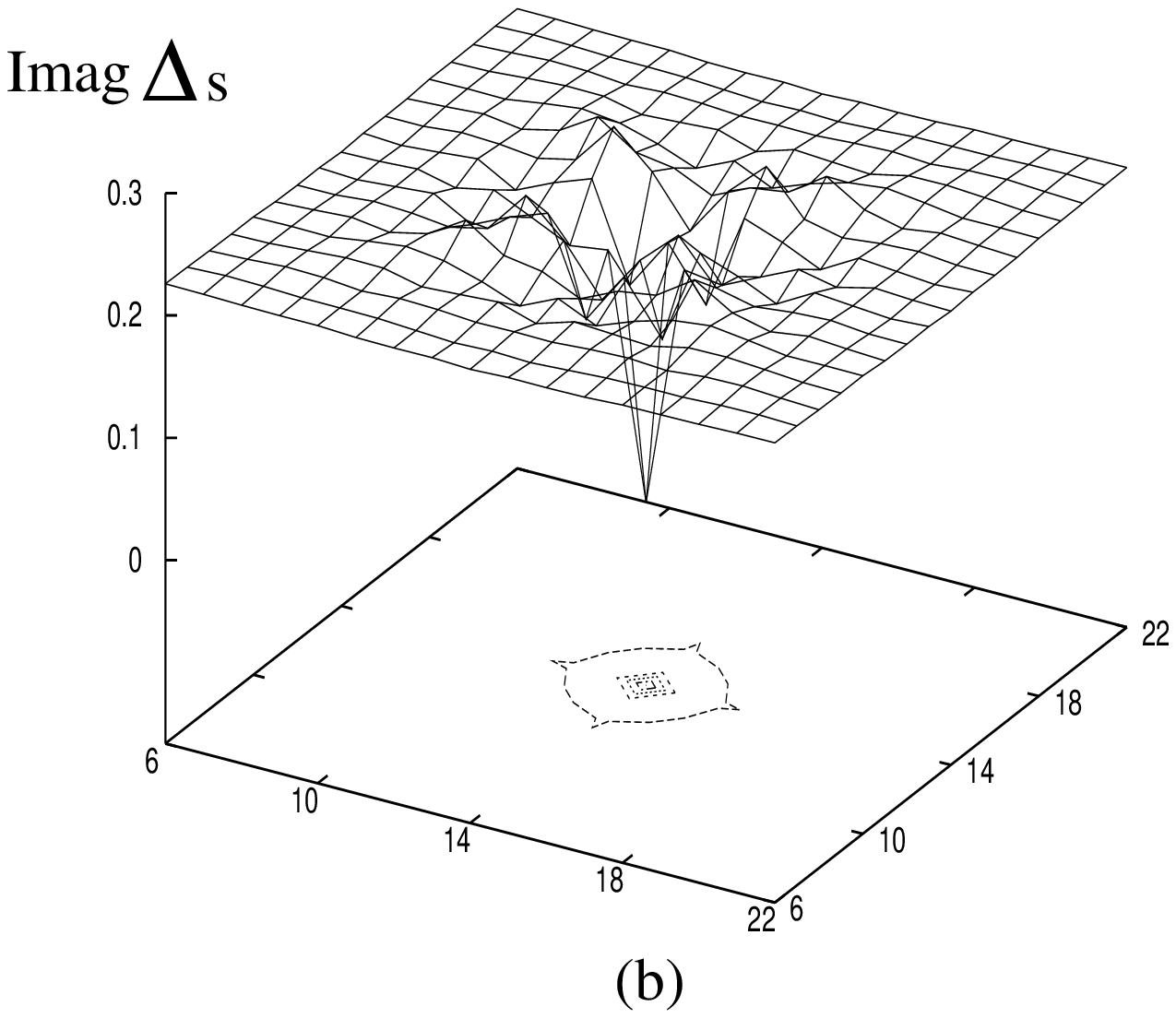,width=6.5cm} \\
   \epsfile{file=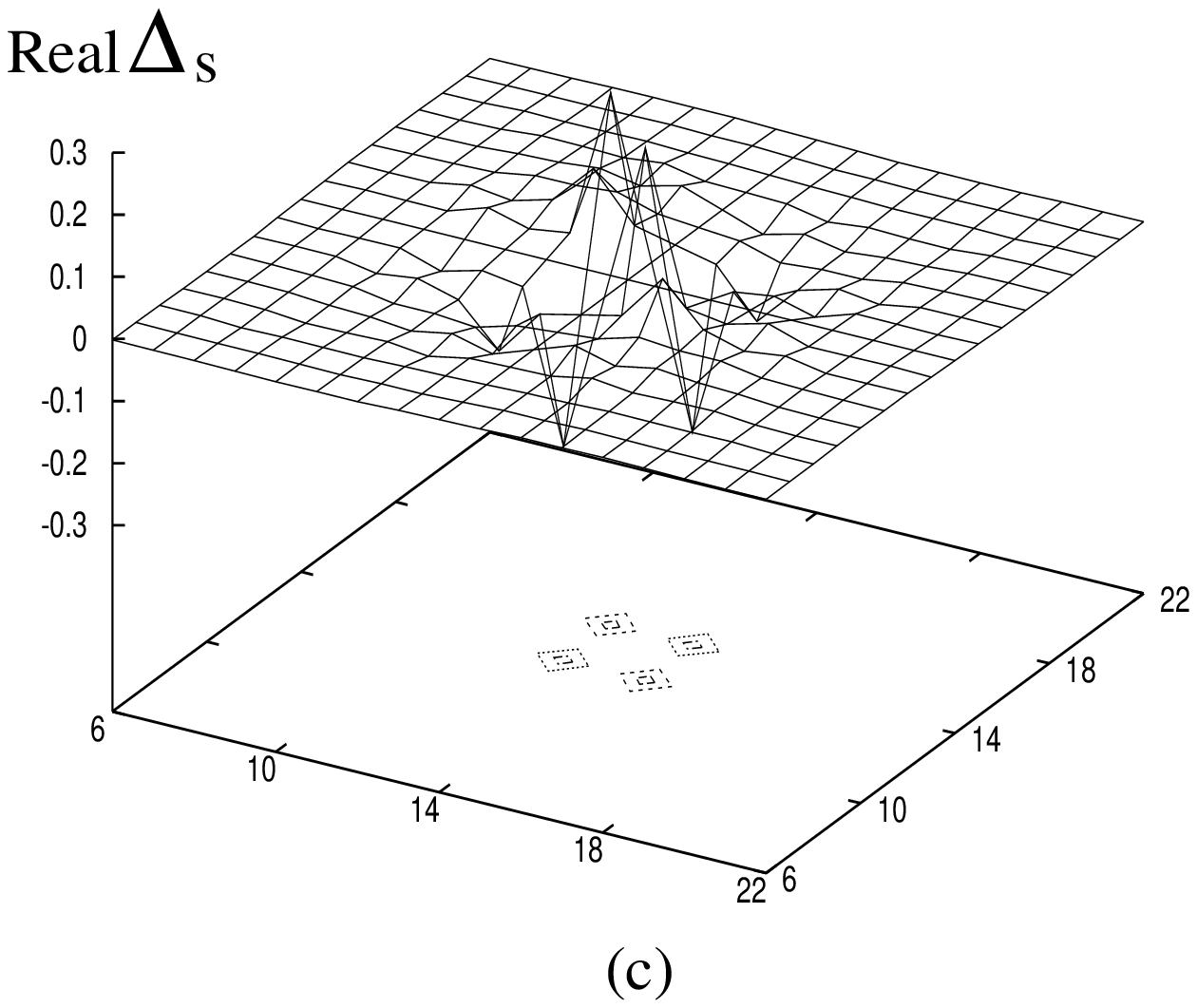,width=6.5cm}   
\caption{ The self-consistent gap function for the 
$d_{xy}+{\rm i}s$ superconductor. Fig.(3-a) and (3-b) shows the 
spatial variation of the $d_{xy}$ and 
$s$-wave components, respectively. Fig.(3-c) is the induced real 
component of the s-wave gap which is absence in the bulk.
}
\label{fig:fig3abc}
\end{figure}
The real s-wave component is induced along the diagonal 
direction and vanishes in the node  directions of the d-wave gap,
and change sign under 90-degree rotation 
as in previous studies~\cite{rf:Franz,rf:Tanaka}, which are 
reflections of the sign change of the d-wave component.
The induced current around the impurity is shown in Fig.4. 
In this case, the induced current  is not circular but 
flows along the diagonal direction. 
\begin{figure}[t]
\begin{center}
   \epsfile{file=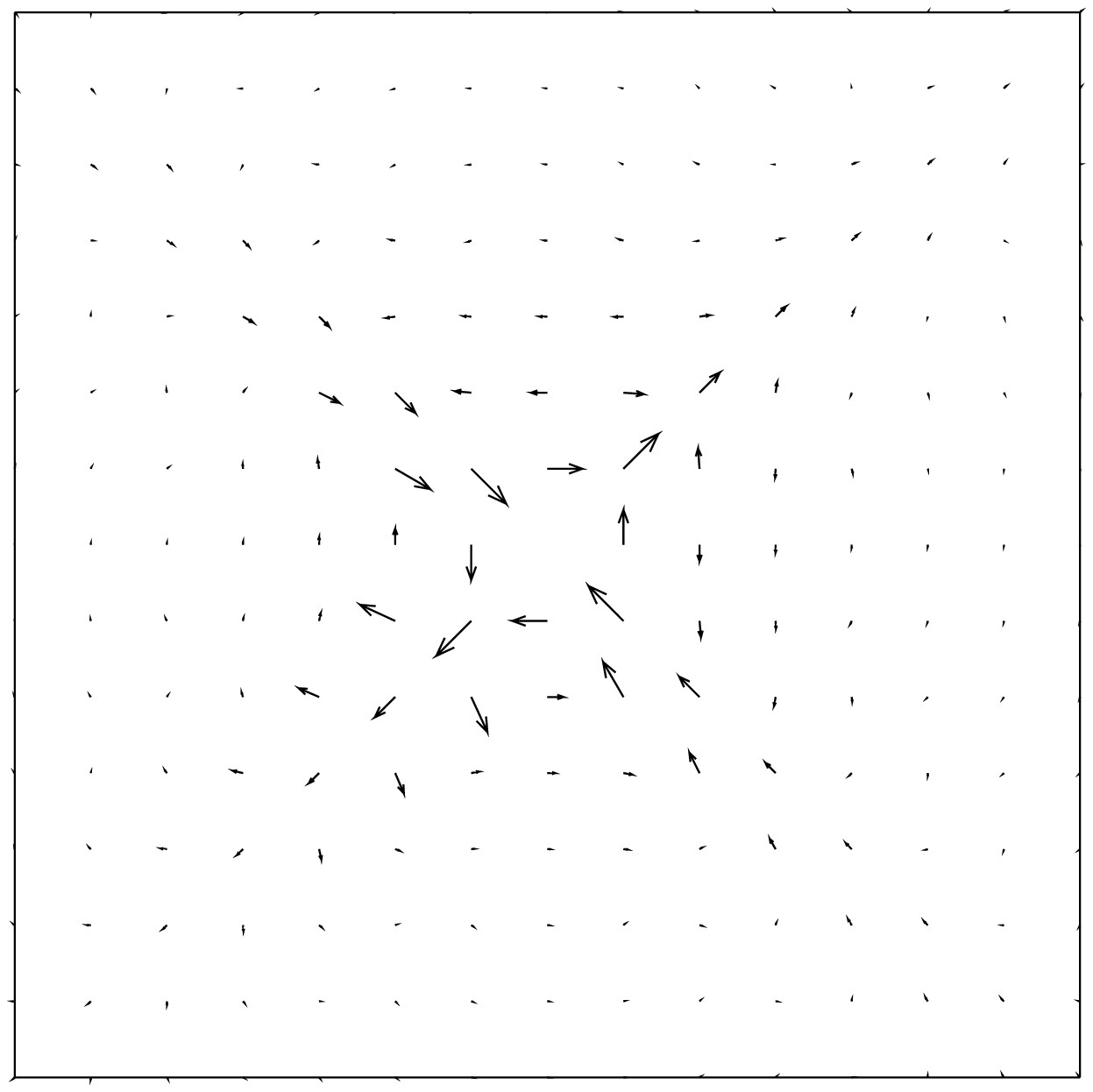,width=5cm}
\end{center}
\caption{ 
Induced current around the impurity in
$d_{xy}+{\rm i}s$ superconductor.
}
\label{fig:fig4}
\end{figure}
The pattern of the current is understood from the Ginzburg-Landau 
theory. \\
\noindent The Ginzburg-Landau free energy is given as, 
\begin{eqnarray}
f&=& \int d^{2}{\rm r}[\sum_{j=d,s}(\alpha_{j}|\Delta_{j}|^{2} 
+\beta_{j}|\Delta_{j}|^{4}+K_{j}|{\bf \Pi }\Delta_{j}|^{2})+
\gamma_{1}|\Delta_{d} |^{2}|\Delta_{s} |^{2}+
\frac{1}{2}\gamma_{2}(\Delta^{\ast 2}{}_{d}\Delta_{s}{}^{2}+
\Delta_{d}{}^{2}\Delta^{\ast 2}{}_{s} ) \nonumber \\
&+&\tilde{K}(\Pi^{\ast}{}_{X}\Delta_{s}\Pi_{X}\Delta^{\ast }{}_{d}
-\Pi^{\ast}{}_{Y}\Delta_{s}\Pi_{Y}\Delta^{\ast }{}_{d}+{\rm c.c})
+g_{s}\delta ({\bf r})|\Delta_{s} |^{2}
+g_{d}\delta ({\bf r})|\Delta_{d} |^{2}], \label{GLfree}
\end{eqnarray}
where $g_{s}$ and $g_{d}$ are the impurity potential for 
s and d-wave respectively and  
$X$ and $Y$ are directed to the diagonal (1,1) and (-1,1) 
respectively and  ${\bf \Pi}=-{\rm i}(\nabla_{X},\nabla_{Y})$.
The current are expressed as,
\begin{eqnarray}
{\bf j}({\bf r}) &=& (-2{\rm e})[
(K_{s}\Delta^{\ast}{}_{s}{\bf \Pi}{}^{\ast} \Delta^{\ast}{}_{s} 
+K_{d}\Delta^{\ast}{}_{d}{\bf \Pi}{}^{\ast} \Delta^{\ast}{}_{d}) \nonumber \\
&+& 
\tilde{K}\{ (\Delta^{\ast}{}_{s}\Pi^{\ast }{}_{X}\Delta_{d}
+\Delta^{\ast}{}_{d}\Pi^{\ast }{}_{X}\Delta_{s})
\hat{{\rm e}}{}_{X}
+{\rm c.c} \} \nonumber \\
&-& \tilde{K}\{ (\Delta^{\ast}{}_{s}\Pi^{\ast }{}_{Y}\Delta_{d}
+\Delta^{\ast}{}_{d}\Pi^{\ast }{}_{Y}\Delta_{s})
\hat{{\rm e}}{}_{Y}
+{\rm c.c} \} ]
\end{eqnarray}
where $\hat{{\rm e}}{}_{X}$ and $\hat{{\rm e}}{}_{Y}$ are the 
unit vectors in the $X-$ and $Y-$ direction respectively.
The parts which come from the 
mixing gradient term ($\tilde{K}$ term ) is important
for the induced current. The d-wave gap are most affected by 
the non-magnetic impurity, and the suppression of the gap occurs 
in the diagonal direction. So the current in the diagonal 
direction is most remarkable.
Due to the sign change of the 
d-wave gap, the direction of the current is reversed 
between $X-$ and $Y-$ direction. If we consider the conservation of 
the current, the overall feature of the induced current 
should be given like Fig.5, which we cannot see 
unfortunately them easily in the numerical data due to the 
their small value.
But it is a natural feature because the superconducting state 
is admixture of the orbital angular momentum of Cooper pair,  
${\rm L}_{z}=2$ and ${\rm L}_{z}=-2$ ( and s-wave component 
${\rm L}_{z}=0$). The direction of the current circle is 
the appearance of the mixture of ${\rm L}_{z}=2$ 
and ${\rm L}_{z}=-2$ 
of the Cooper pair. \\
\begin{figure}[t]
\begin{center}
   \epsfile{file=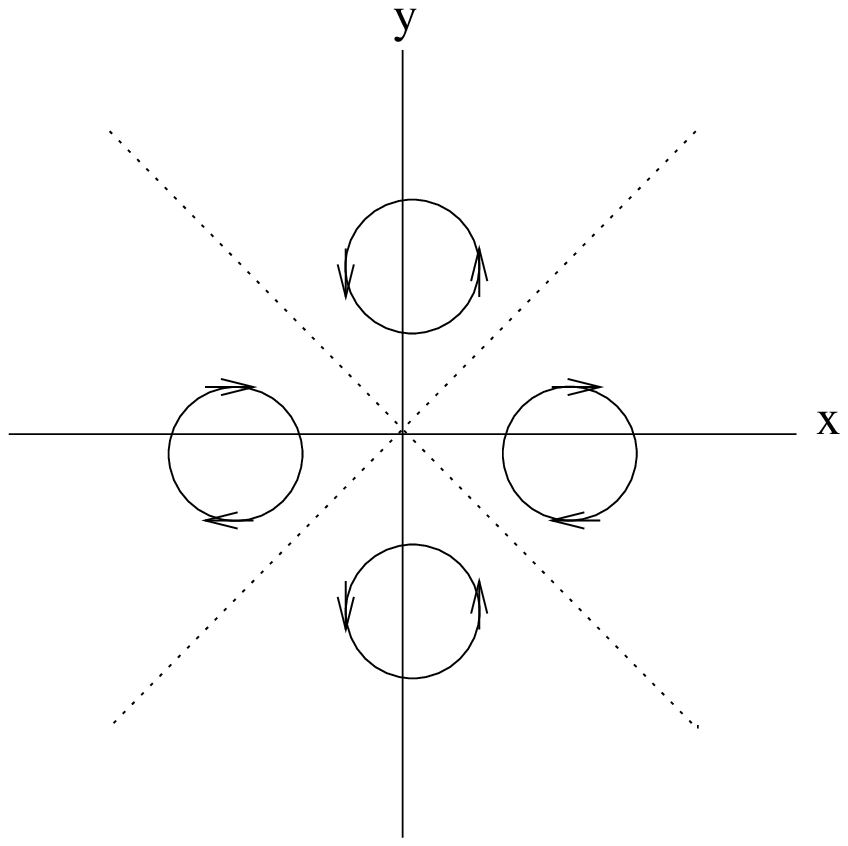,width=5cm}
\end{center}
\caption{ 
Schematic figure of the induced current around the 
impurity for $d_{xy}+{\rm i}s$ superconductor.
}
\label{fig:fig5}
\end{figure}
The induced current is clearly 
distinguished from that of the chiral $p_{x}+{\rm i}p_{y}$ one.
In chiral $p_{x}+{\rm i}p_{y}$ case, the induced current forms the 
magnetic moment around the impurity, but 
in the $d_{xy}+{\rm i}s$ case, 
the induced current is arranged like a quadrapole, which 
is just the reflection of the angular momentum of $d_{xy}$ 
component. We discuss the magnetic field generated by this 
type of Cooper pair ($d+{\rm i}s$) in Appendix 2. 
The $d_{x^{2}-y^{2}}+{\rm i}s$, state 
which are often discussed 
in the context of the high temperature superconductor, will give 
the same type of the induced current with 90 degree rotation about 
the $z$ axis. \\
In this section we consider the superconductor which 
violates the ${\cal T}$ symmetry in the bulk.
It is possible that the ${\cal T}$-violating 
superconducting state in  realized  
locally around the impurity~\cite{rf:Balatsky}.
The induced current pattern will be 
the same as that of the corresponding 
${\cal T}$-violating bulk one. So if we can detect the 
magnetic field pattern around the impurity, we can 
distinguish the symmetry of the Cooper pair. 
But it is difficult to produce the locally ${\cal T}$-violating 
superconducting state only by the non-magnetic or 
magnetic impurity without spin-orbit interaction. We can understand  
the reason in the  GL treatment. Based on 
GL free energy (\ref{GLfree}) and we assume the bulk state 
is the real d-wave components only. 
The spatial variation of the 
order parameter  induces the s-wave component 
near the impurity.
If the impurity effect is 
not so strong, we expand the GL equation by the impurity 
potential. We can easily see the first order of the 
impurity potential, only the real part of the s-wave and 
d-wave component coupled to the impurity potential.  
So the phase of the induced s-wave is real with in the 
first order of the impurity potential and does not 
violate the ${\cal T}$ symmetry. \\
\section{Magnetic impurity in p$_{x}$+{\rm i}p$_{y}$ 
superconductor }
In this section we discuss the magnetic impurity effects in 
chiral state, p${}_{x}$+{\rm i}p${}_{y}$ superconductor. 
Here we limit our considerations to a classical spin, $S >>1$. 
In the singlet case we can take the direction 
of the impurity spin parallel to the z-axis by suitable unitary 
transformation.  In the case of the triplet superconductor, we cannot  
perform the same operation due to the lack of the spin rotation 
symmetry. But we take the direction of the magnetic 
impurity in the z direction due to the physical condition 
as we explains below.
The spontaneous circular current will flow around the magnetic 
impurity and the magnetic field generated by the 
induced current will 
fix the direction of impurity spin to the z-axis.
So we take the  magnetic impurity as a classical 
Ising type and neglect the transversal components of the 
impurity spin. \\
The Bogoliubov-de Gennes equation with the simple Ising-like 
magnetic impurity is given like below,
\begin{eqnarray} \begin{array}{l}
{\rm h_{0}}({\bf r}) u_{\rm i \sigma}({\bf r}) 
+\sigma {\rm JS_{z}}\delta ({\bf r})u_{\rm i \sigma}({\bf r})
+ \sum_{\sigma'} \int d{\bf r^{'}} \hat{\Delta }_{\sigma,\sigma'}
({\bf r},{\bf r^{'}}) v_{\rm i -\sigma'}({\bf r^{'}}) 
= {\rm E_{\rm i}}u_{\rm i \sigma}({\bf r}) \\
-{\rm h_{0}}({\bf r})v_{\rm i -\sigma}({\bf r}) 
+\sigma {\rm JS_{z}}\delta ({\bf r})v_{\rm i -\sigma}({\bf r})
- \sum_{\sigma'}\int d{\bf r^{'}} \hat{\Delta^{\ast }}_{\sigma \sigma'}
({\bf r},{\bf r^{'}})u{}_{i \sigma'}({\bf r^{'}}) 
= {\rm E_{\rm i}}v{}_{\rm i -\sigma}({\bf r}), \label{Bogol}  
\end{array} \end{eqnarray}
where ${\rm J}$ is the  exchange energy and ${\rm S_{z}}$ is 
the impurity spin.
The BdG equation is decoupled by $(u{}_{\uparrow}({\bf r}),
v{}_{\downarrow}({\bf r}) )$ and 
$(u{}_{\downarrow}({\bf r}),v{}_{ \uparrow}({\bf r}) )$ 
due to the absence of the transversal component of the 
imputiry spin. 
By the solution of 
$(u{}_{\uparrow}({\bf r}),v{}_{\downarrow}({\bf r}) )$ 
with the eigenvalue ${\rm E}_{\rm i}$, we can construct 
the solution of the pair 
$(u{}_{\downarrow}({\bf r}),v{}_{ \uparrow}({\bf r}) )$ 
with the eigenvalue $-{\rm E}_{\rm i}$ like 
$(v^{\ast }{}_{\downarrow }({\bf r}),
u^{\ast }{}_{\uparrow}({\bf r}))$ and this is an one to one 
correspondence. First, we neglect the spatial dependence of 
the gap function, and fix
$\Delta (k)=\Delta_{0}(k_{x}+{\rm i}k_{y})$.
The BdG equation (\ref{Bogol}) has two types of the solutions, 
one is the continuum state and the other is the bound state 
at the impurity. The bound state is degenerate and  is 
explicitly given by,
\begin{equation}
\left (
\begin{array}{c}
u_{{\rm B}\uparrow }({\bf r}) \\
v_{{\rm B}\downarrow }({\bf r})
\end{array}
\right )
=
\left (
\begin{array}{c}
\frac{2{\rm I}_{0}{\rm N}_{0}\epsilon_{-}}
{\sqrt{\Delta_{0}{}^{2}-\epsilon_{-}{}^{2}}}f_{1}(k_{F}r) \\
-\frac{2{\rm I}_{0}{\rm N}_{0}\Delta_{0}}
{\sqrt{\Delta_{0}{}^{2}-\epsilon_{-}{}^{2}}}
{\rm i}f_{2}(k_{F}r){\rm e}^{-{\rm i}\theta}
\end{array}
\right ) 
{\rm or } 
\left (
\begin{array}{c}
-\frac{2{\rm I}_{0}{\rm N}_{0}\Delta_{0}}
{\sqrt{\Delta_{0}{}^{2}-\epsilon_{-}{}^{2}}}
{\rm i}f_{2}(k_{F}r){\rm e}^{{\rm i}\theta} \\
-\frac{2{\rm I}_{0}{\rm N}_{0}\epsilon_{-}}
{\sqrt{\Delta_{0}{}^{2}-\epsilon_{-}{}^{2}}}f_{1}(k_{F}r) 
\end{array}
\right ) \label{bound2}
\end{equation}
where ${\rm N}_{0}$ is the density of state at the Fermi 
level and ${\rm c}=\pi {\rm N}_{0}{\rm JS_{z}}$ and 
${\rm I}_{0}{}^{2}
=\frac{2\Delta_{0}}{\pi {\rm N}_{0}}
(\frac{{\rm c}^{2}}{1+{\rm c}^{2}})^{\frac{3}{2}}$, and 
the functions $f_{1}(k_{F}r)$ and $f_{2}(k_{F}r)$,
where $k_{F}$ is the Fermi wave number, are 
approximately given for the distance $r>>1/k_{F}$, as  
\begin{equation} \begin{array}{l} \displaystyle
f_{1}(k_{F}r) \approx \frac{\pi}{2}\sqrt{\frac{2}{\pi k_{F}r}}
{\rm cos}(k_{F}r-\frac{\pi}{4})
{\rm e}^{-\frac{\sqrt{\Delta_{0}{}^{2}-\epsilon^{2} }}
{v_{F}}r} 
\\ \\ \displaystyle
f_{2}(k_{F}r) \approx \frac{\pi}{2}\sqrt{\frac{2}{\pi k_{F}r}}
{\rm cos}(k_{F}r-\frac{3\pi}{4})
{\rm e}^{-\frac{\sqrt{\Delta_{0}{}^{2}-\epsilon^{2} }}
{v_{F}}r}.
\\
\end{array} \end{equation}
$\epsilon_{-}$ is an 
eigenvalue of the bound state and is 
given as  
\begin{equation}
\epsilon_{-} = -\frac{\Delta_{0}}{\sqrt{1+{\rm c}^{2}}}.
\end{equation} 
We can see the mixing of the angular momentum of the 
wave functions in (\ref{bound2}). Here the above eigenvalue 
and the bound states are derived by the assumption of the
electron-hole symmetry and the infinite band width of the 
conduction electron.
We can easily see that for the electron-like state the
particle wave function couples to the hole wave function of an
angular momentum reduced by 1. For the hole-like states it is just
the opposite way around. This property is responsible for the fact that
these states carry a circular current.
The solution of the bound state with an eigenvalue  
$\epsilon_{+}=\frac{\Delta_{0}}{\sqrt{1+{\rm c}^{2}}}$ is 
given as, 
$(u{}_{{\rm B}\downarrow}({\bf r}),v{}_{{\rm B} \uparrow}({\bf r}) )$ 
=$(v^{\ast }{}_{{\rm B}\downarrow }({\bf r}),
u^{\ast }{}_{{\rm B}\uparrow}({\bf r}))$, as we said. 
The spectral density, 
$A_{\sigma}({\rm r}=0,\omega )
=Z^{(+)}{}_{\sigma}(0)\delta(\omega -\epsilon )
+Z^{(-)}{}_{\sigma}(0)\delta(\omega +\epsilon )$, 
 at the impurity site due to the 
bound state is given like, 
$Z^{(+)}{}_{\downarrow} = Z^{(-)}{}_{\uparrow} = 
\frac{2\pi {\rm N}_{0}\Delta_{0}{\rm c}}
{(1+{\rm c}^{2})^{\frac{3}{2}}}$, and  
$Z^{(-)}{}_{\downarrow} = Z^{(+)}{}_{\uparrow} = 0$. \\
The current composed by the bound state is given below,
\begin{eqnarray}
j_{\theta }{}^{B} &=& 
\frac{4{\rm e}\Delta_{0}{\rm N}_{0}|{\rm c}|}
{\pi m (1+{\rm c}^{2})^{\frac{1}{2}}}
\frac{f_{2}(k_{F}r)^{2}}{r}
\end{eqnarray}
In contrast to the non-magnetic impurity~\cite{rf:Okuno}, 
there is finite electron-like  
contribution of the current at the absolute zero temperature.
This is due to the fact that the attractive nature 
of the exchange impurity potential (${\rm J}>0$) for the 
down spin electrons.  \\
We can also calculate the contribution of the continuum state 
to the induced current as in the non-magnetic 
impurity case~\cite{rf:Okuno}. In the unitary limit 
(${\rm c}\rightarrow \infty $), the current from the 
continuum is given by,
\begin{eqnarray}
j_{\theta }{}^{c} &\sim& -\frac{{\rm e}\Delta_{0}}
{2{\rm m}\pi^{2}v_{F}}(2\pi-1)
\frac{{\rm cos}^{2}(k_{F}r-\frac{3}{4})}{r^{2}}
(1+{\rm e}^{\frac{-2\Delta_{0}}{v_{F}}r}),
\end{eqnarray}
where $v_{F}$ is the Fermi velocity. The power-law 
decay $1/r^{2}$ is because of the fact that  
the current is carried by 
the quasiparticle belonging to the continuum spectrum.
\\
Basically, the feature and the mechanism of 
the induced current is not different from the non-magnetic case 
and the contribution of the continuum spectrum is important.
The direction of the current composed by the bound state is 
opposite to that of the continuum one. \\
We can see the induced circular current also flows around 
the magnetic impurity and it does not contradict the 
assumption  we first set up. \\
Due to the spin degree of freedoms 
magnetic impurities 
cause another effect 
to the superconducting state. 
In the s-wave superconductor,  there 
is a zero temperature first-order phase transition 
depending on the magnitude of the impurity exchange potential
${\rm J}$. 
The total spin of the electron change from 
spin unpolarized state, $<s_{z}>=0$, to the polarized one 
$<s_{z}>=-\frac{1}{2}$, which was noticed 
by Sakurai~\cite{rf:Sakurai}. 
The physical picture 
of the phase transition is that the electron is captured 
by the magnetic imputiry due to the large enough 
exchange energy, ${\rm J}$. So there is a competition 
between the pairing-condensation energy and the magnetic 
impurity interaction. 
The feature of the transition can 
be understood from the level crossing of the bound 
state~\cite{rf:Sakurai,rf:Nakajima}. 
We express the solution of the BdG equation of  
the pair 
$(u{}_{{\rm n}\uparrow}({\bf r}),
v{}_{{\rm n}\downarrow}({\bf r}) )$ 
as $\gamma_{n}$
and as $\gamma^{'}{}_{m}$ for the solution of the 
pair $(u{}_{{\rm m}\downarrow}({\bf r}),
v{}_{{\rm m}\uparrow}({\bf r}) )$.
We set the ${\rm n}>0$ for the eigenvalue 
${\rm E}{}_{{\rm n}} > 0$ and ${\rm n}<0$ for 
${\rm E}{}_{{\rm n}} < 0$. 
The difference of the number of up spin 
electrons and that of down spin electrons, ${\rm M}_{z}$
is expressed as,
\begin{eqnarray}
{\rm M}_{z} &=& \sum_{{\rm n}<0} 1 - \sum_{k} 1 
+\sum_{{\rm n}>0} \gamma^{\dagger}{}_{n}\gamma_{n}-
\sum_{{\rm m}>0} 
\gamma^{'\dagger}{}_{{\rm m}}\gamma^{'}{}_{{\rm m}}
\end{eqnarray}
where the summation $\sum_{k}$ runs over 
the wave number space.
The spin of the ground state is given by,  
$\sum_{{\rm n}<0} 1 - \sum_{k} 1 $,  and 
changes its value when the number of the negative energy 
states are decreased or increased, namely it is 
the level crossing across 
the Fermi level.
We can see the energy level of the bound state for 
the $p_{x}+{\rm i}p_{y}$ superconductor is 
$\epsilon_{B}=-\Delta_{0}/\sqrt{1+{\rm c}^{2}}$. 
It does not cross the Fermi level however large the 
exchange energy of the impurity is. So it can be concluded 
that a phase transition, as found in an s-wave superconductor,  
is absent in this case. This conclusion is also 
applicable for the d-wave superconductor and was already 
pointed out by Salkola~\cite{rf:Salkola}.  
However, we used various assumptions like 
electron-hole symmetry and impurity s-wave scattering. 
These are  rather ideal conditions for 
the actual magnetic impurity systems. 
If we give up these restrictions, the 
situation is changed. For example, if we allow the electron-hole 
asymmetry like 
${\rm N}(\xi )={\rm N}{}_{0}+{\rm N}^{'}\xi $, where  
${\rm N}(\xi )$ is the density of state and ${\rm N}^{'}$ 
is the introduced phenomenological parameter for 
the electron-hole asymmetry. Then the level of the bound state 
is given as, 
\begin{eqnarray}
\epsilon{}_{\pm} = \pm \frac{(1-{\rm N}^{'}{\rm D}{\rm JS_{z}})}
{\sqrt{(\pi {\rm N}_{0}{\rm JS_{z}})^{2}
+(1-2{\rm N}^{'}{\rm D}{\rm JS_{z}})^{2}}}\Delta_{0}
\end{eqnarray}
where ${\rm D}$ is the half of the band width.  We can see 
that the level crossing can occur at the critical strength, 
${\rm J_{c}S_{z}}=\frac{1}{{\rm N}^{'}{\rm D}}$. \\
We can also 
see that p-wave scattering at the impurity 
can also  cause the phase transition to the spin polarized 
state. For example, we take the magnetic impurity potential 
as,
\begin{eqnarray}
\hat{{\rm J}}({\bf k},{\bf k^{'}}) &=& 
{\rm J}_{s}{\bf \sigma_{z}}+
{\rm J}_{p}({\rm e}^{{\rm i}(\phi-\phi^{'})}+
{\rm e}^{-{\rm i}(\phi-\phi^{'})}){\bf \sigma_{z}} 
\label{magpoten}
\end{eqnarray} 
where ${\rm J}_{s}$ and 
${\rm J}_{p}$ is the s-wave and p-wave scattering 
term, respectively, and $\phi$ and $\phi^{'}$ is the 
angle of ${\bf k}$ and ${\bf k^{'}}$ respectively. 
In this case the energy levels of the 
bound states solution for the pair 
$(u{}_{{\rm n}\uparrow}({\bf r}),
v{}_{{\rm n}\downarrow}({\bf r}) )$ 
are given like below (see Appendix3),
\begin{subeqnarray}
\epsilon_{1} &=& -\frac{(1-{\rm cd})\Delta_{0}}
{\sqrt{({\rm c+d})^{2}+(1-{\rm cd})^{2}}} \\
\epsilon_{2} &=& -\frac{\Delta_{0}}{\sqrt{1+{\rm d}^{2}}} 
\label{pwbound}
\end{subeqnarray}
where ${\rm c}=\pi{\rm N}_{0}{\rm J}_{s}{\rm S}$ and 
${\rm d}=\pi{\rm N}_{0}{\rm J}_{p}{\rm S}$. 
We can see from (\ref{pwbound}a) that the 
impurity spin can be compensated when ${\rm c} > 1/{\rm d}$.
Another bound state is also accompanied by introducing the 
p-wave scattering (\ref{pwbound}b). 
 We take another impurity potential like below,
\begin{eqnarray}
{\hat {\rm J}}(k,k^{'}) &=& 
{\rm U}_{s}+
{\rm J}_{p}({\rm e}^{{\rm i}(\phi-\phi^{'})}+
{\rm e}^{-{\rm i}(\phi-\phi^{'})}){\bf \sigma_{z}} \label{magpoten2}
\end{eqnarray}
where ${\rm U}$ is the s-wave scalar potential. In this case, 
the levels of the bound state is given like,
\begin{subeqnarray}
\epsilon_{1} &=& -\frac{1-{\rm cd}}
{\sqrt{({\rm c+d})^{2}+(1-{\rm cd})^{2}}}\Delta_{0} \\
\epsilon_{2} &=& -\frac{\Delta_{0}}{\sqrt{1+{\rm d}^{2}}} \\
\epsilon_{3} &=& \pm \frac{1+{\rm cd}}
{\sqrt{({\rm c-d})^{2}+(1+{\rm cd})^{2}}}\Delta_{0} 
\label{pwbound2}
\end{subeqnarray} 
where ${\rm c}=\pi{\rm N}_{0}{\rm U}_{s}$ and in 
(\ref{pwbound2}c) we take plus sign  for 
$({\rm c} > {\rm d})$ and minus sign for $({\rm c} < {\rm d})$.
The form of (\ref{pwbound2}a) can also be possible for the 
level crossing.  In the unitary limit for the 
scalar potential ${\rm c}\rightarrow \infty$, the level crossing 
occurs at infinitesimal p-wave magnetic potential. 
The bound state energy level with 
the unitary limit in s-wave scattering, 
${\rm c}\rightarrow \infty$, is 
$\pm {\rm d}\Delta_{0}/\sqrt{1+{\rm d}^{2}}$ and does not 
coincide with the zero energy level. 
The above scenario is applicable to 
the d-wave superconductor with d-wave scattering from the 
impurity.  The 
p-wave and d-wave scattering is 
given naturally from the nearest neighbor 
exchange interaction of the impurity like 
$J_{0}\sum_{\tau}S_{0}\sigma_{0+\tau}$~\cite{rf:Poilblanc,
rf:Tsuchiura}, where summation $\tau$ is taken 
at the nearest neighbors  of the impurity site $0$  
and  $S_{0}$ is the impurity spin and $\sigma $ the electron 
spin. Basically, the impurity resonance peak at the zero energy level 
in unitary limit splits by many factors like electron-hole 
asymmetry and  p-wave scattering. 
The resonance peak splitting of the DOS in 
Tsuchiura~\cite{rf:Tsuchiura}
 may be related to the anisotropic scattering from 
the impurity. The anisotropic scattering from the impurity 
is not negligible in actual transition metal systems like 
the cuprate superconductors.
So actually there is also a first order phase 
transition of the ground state 
in this p-wave case as in the conventional s-wave superconductor.
But the critical strength of the impurity exchange potential 
${\rm J_{c}}$ is not determined only by the competition 
of the condensation energy and the exchange energy as in 
the s-wave case but various factors should be considered. \\
We carry out the numerical calculation of the BdG 
equation as in the non-magnetic impurity.   
In Fig.6 we show the spatial dependence of the gap function.
Basic feature is not different from that of the non-magnetic 
impurity case. The magnetic impurity induces the 
spatial dependence of the spin density.
\begin{figure}[t]
   \epsfile{file=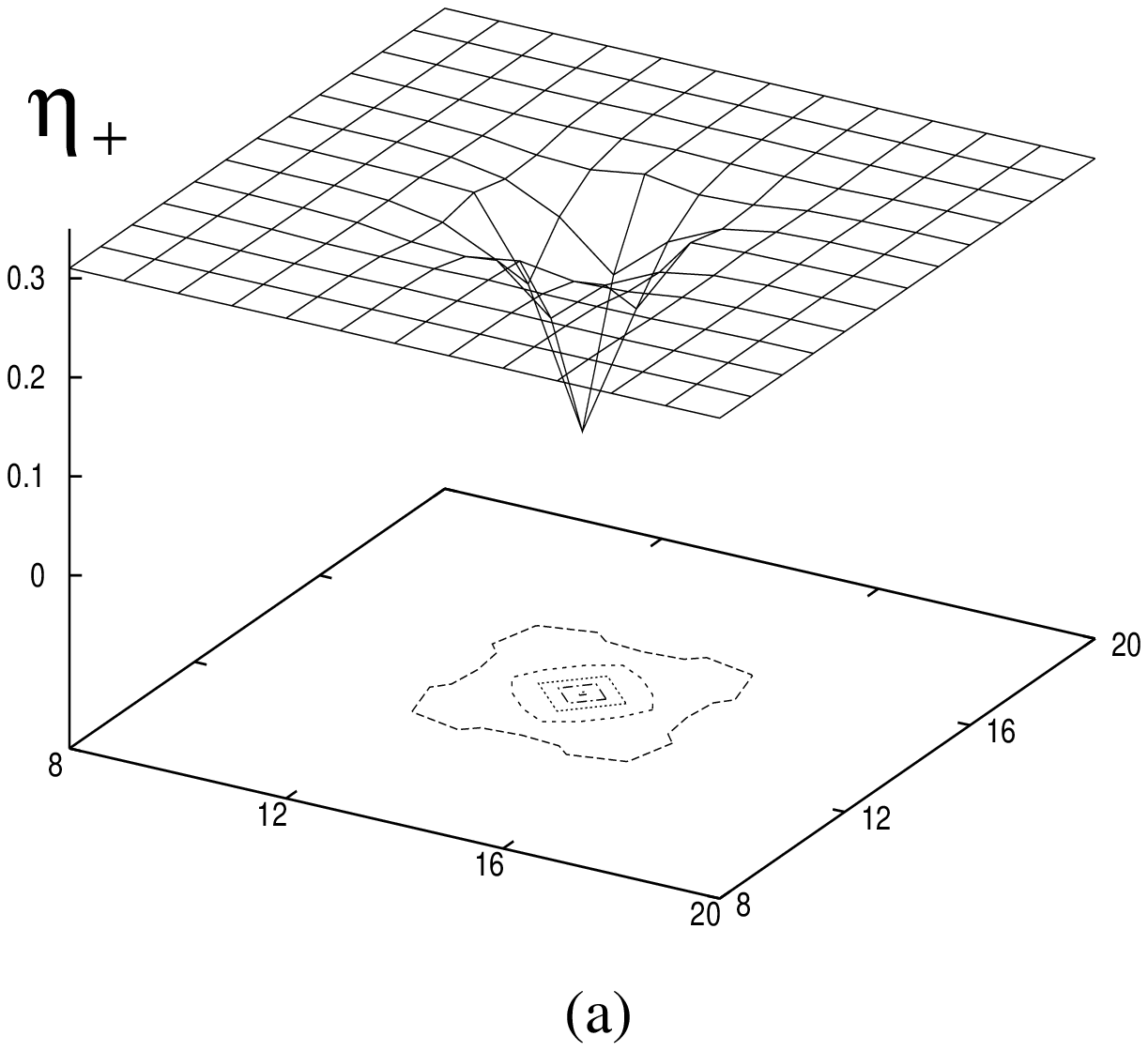,width=6.5cm}
   \epsfile{file=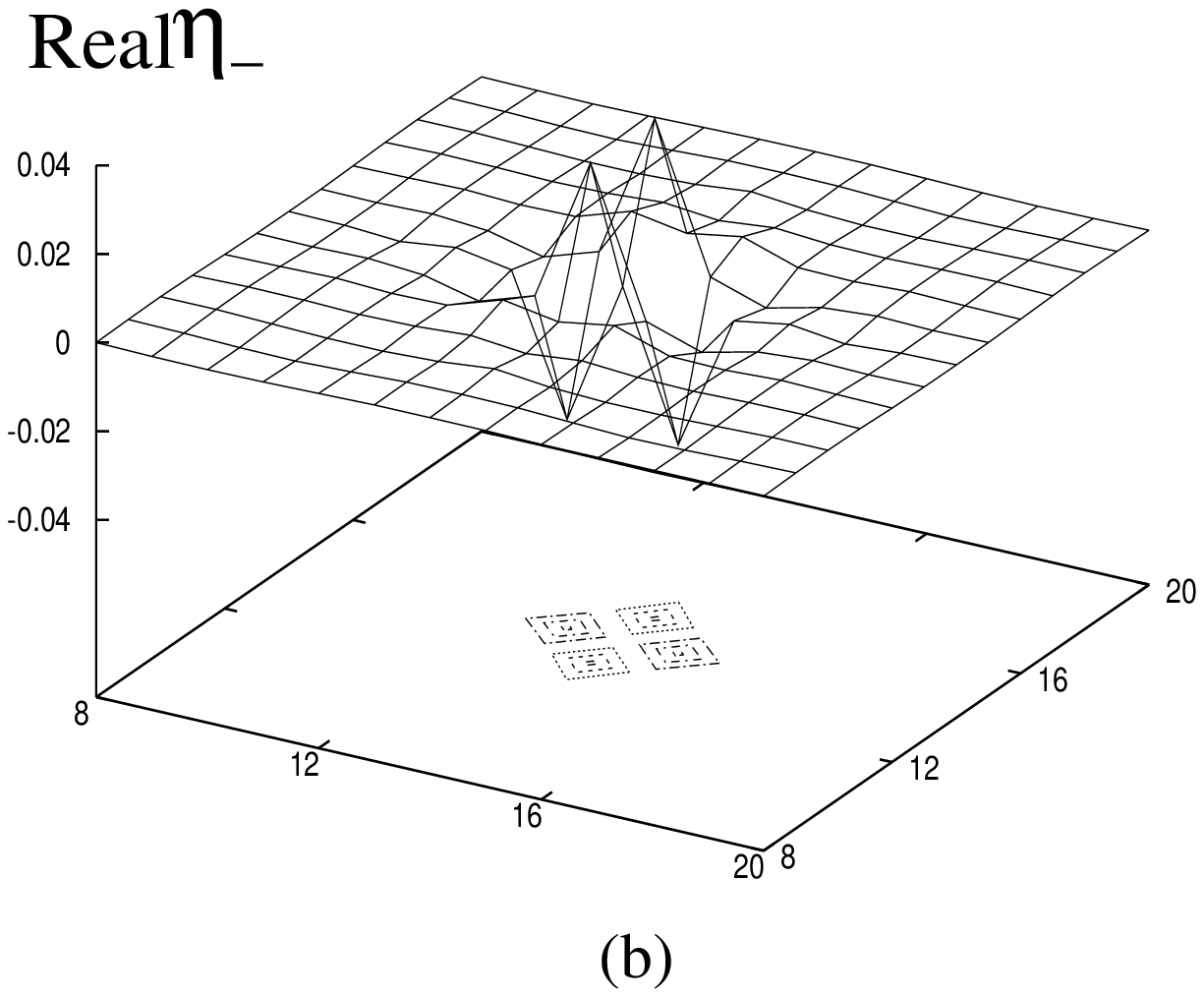,width=6.5cm} \\
   \epsfile{file=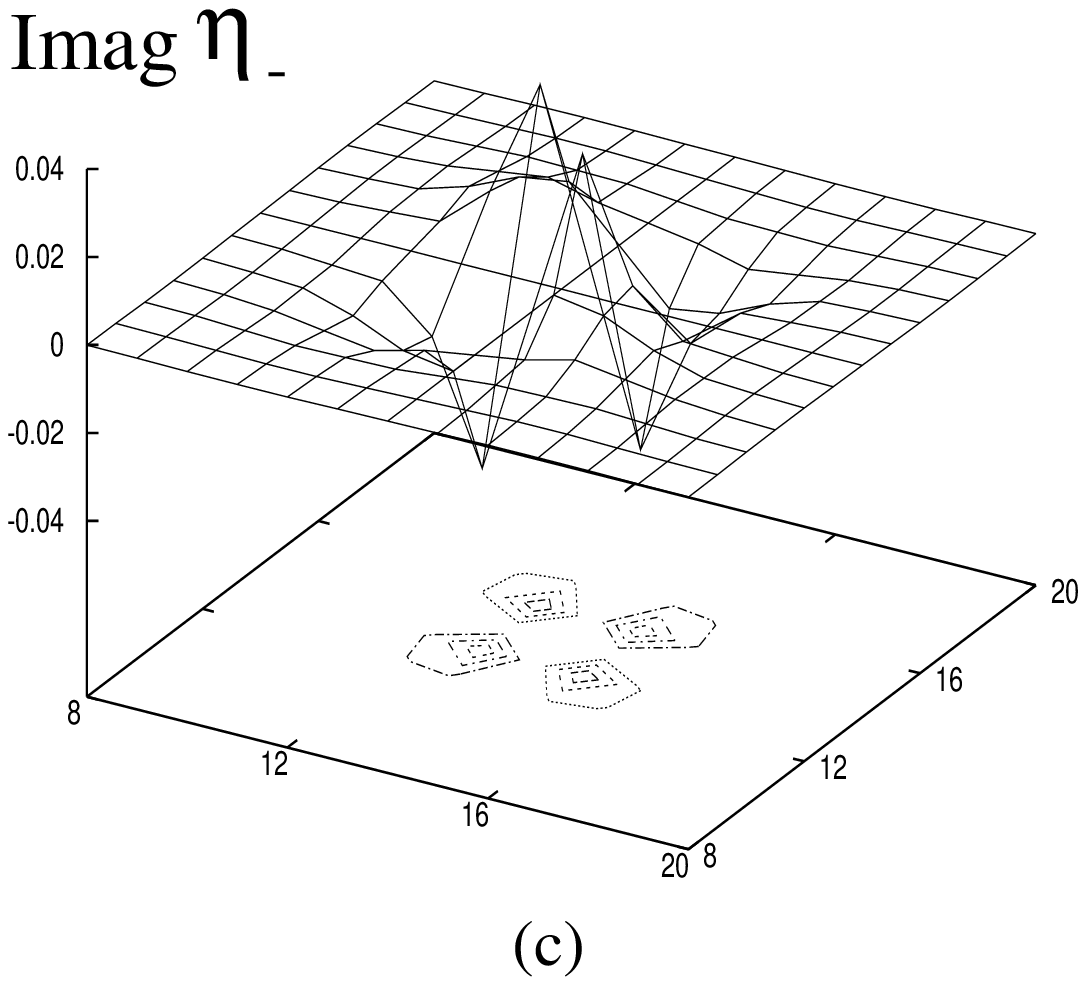,width=6.5cm}   
\caption{ 
The self-consistent gap function for the 
$p_{x}+{\rm i}p_{y}$ superconductor with the magnetic impurity. 
The calculation is done with the parameter $\mu = -1.5t$ and 
${\rm JS}=5t$ on 27$\times $27 lattice. Fig.(6-a) is dominant 
$p_{x}+{\rm i}p_{y}$ component and (6-b), (6-c) is the 
admixture real and imaginary $p_{x}-{\rm i}p_{y}$ component, 
respectively.
}
\label{fig:fig6abc}
\end{figure}
In Fig.7 we show the spin density around the magnetic impurity. 
The peak $<s({\bf r}=0)>$ is cut off in order to illustrate 
the fine details.
In Fig.(7-a) the impurity potential ${\rm JS}=5t$ 
and is not the electron spin-unpolarized state ($<s_{z}>=0$). 
On the other hand, 
(7-b) is the spin-polarized state ($<s_{z}>=-\frac{1}{2}$) 
with ${\rm JS}=10t$.  
\begin{figure}[t]
\begin{center}
   \epsfile{file=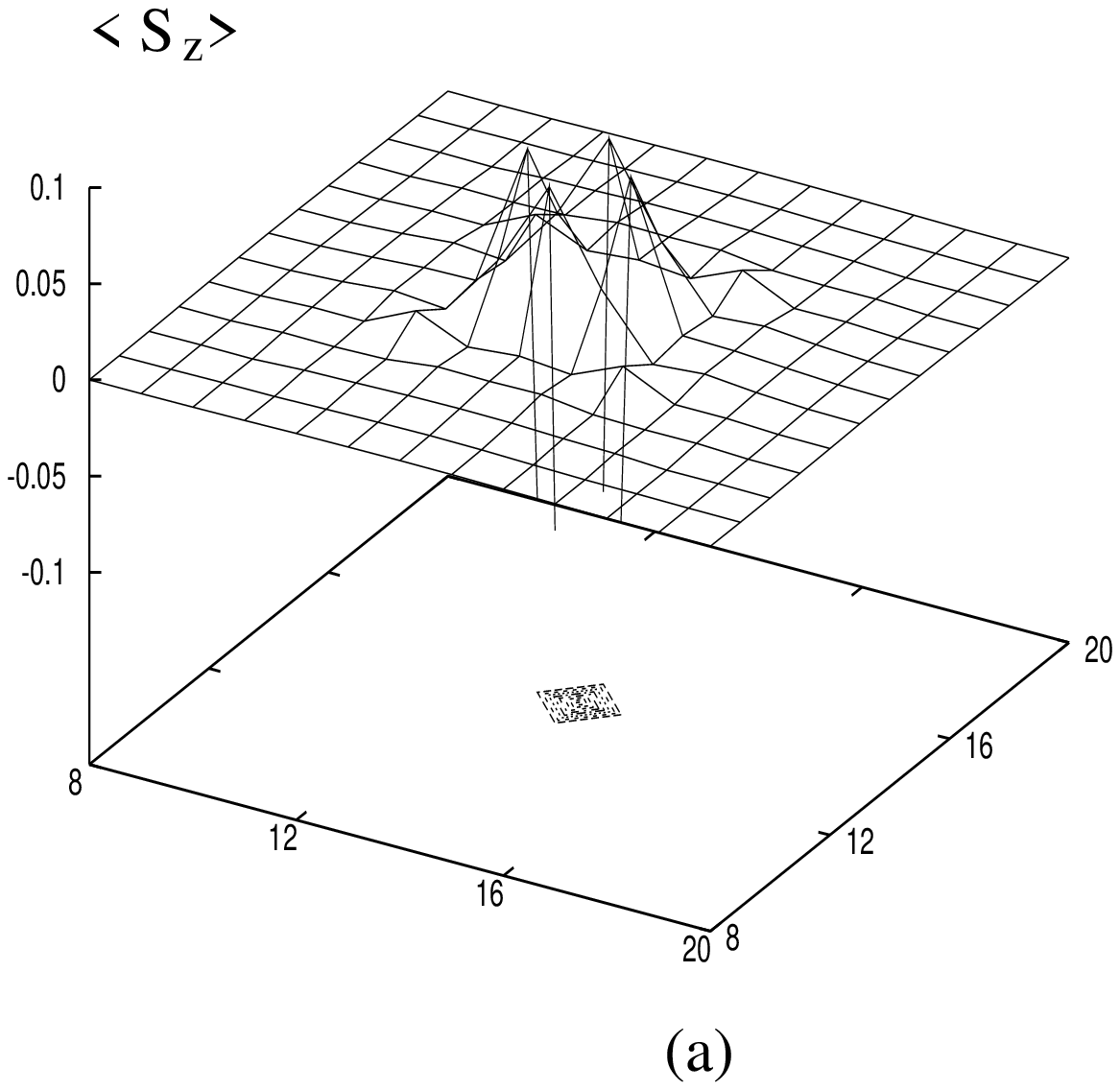,width=6.5cm}
   \epsfile{file=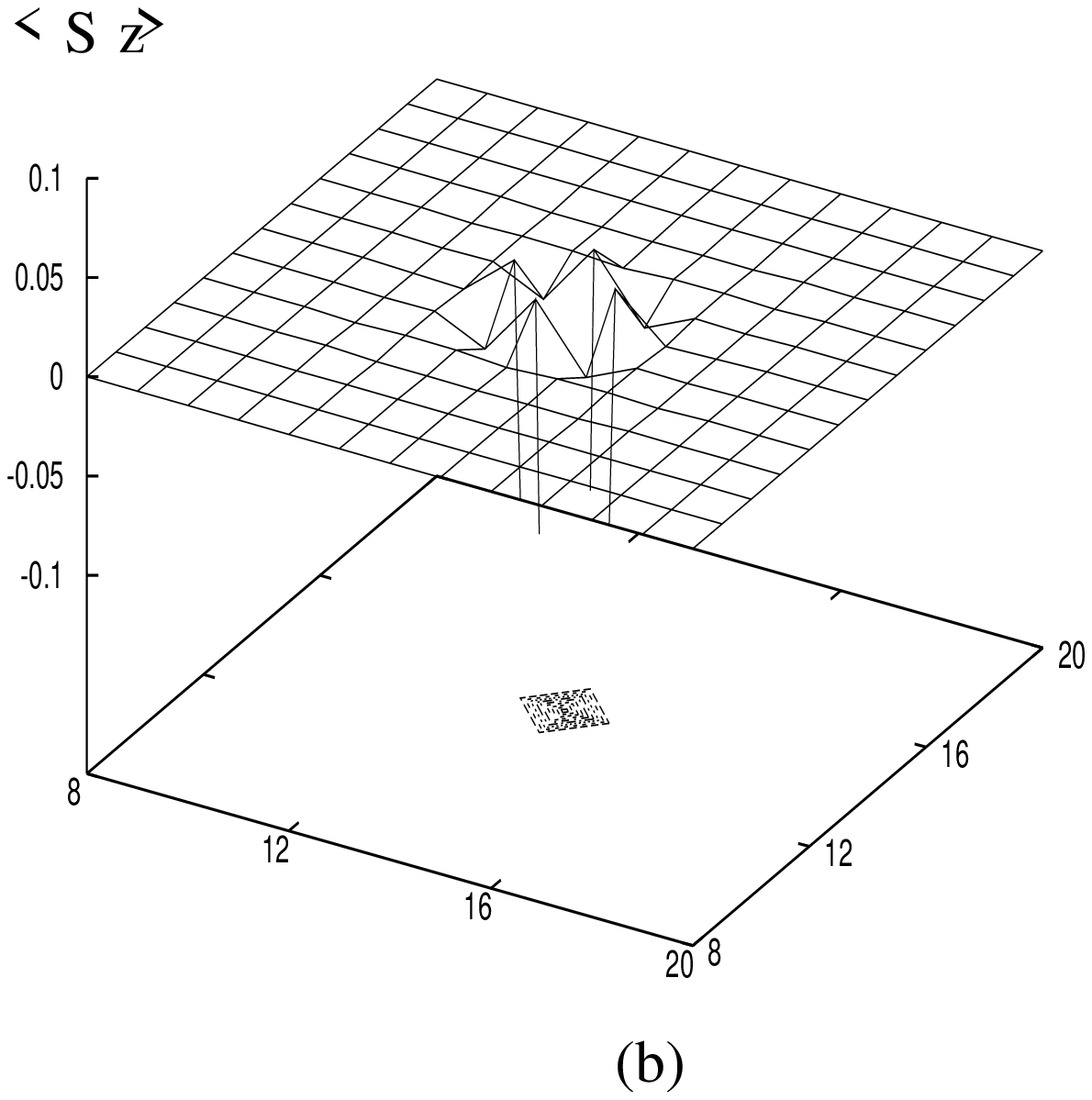,width=6.5cm}
\end{center}
\caption{ 
Spin density around the magnetic impurity. 
The strength of the magnetic impurity ${\rm JS}$ is 5t and 
8t for (a), and (b) respectively. Total spin moment is 
(a) $<s_{z}>=0$ and (b) $<s_{z}>=-\frac{1}{2}$. 
}
\label{fig:fig7}
\end{figure}
There is a compensating spin-density cloud of opposite sign in 
the neighborhood of the magnetic moment in order to 
screen the  spin polarization as  in the d-wave 
case.~\cite{rf:Salkola} On the other hand, for the spin-polarized 
state, the compensating spin-density cloud is much smaller 
than that of unpolarized one due to the imbalance of the 
spin up and down electron. Just at the impurity 
site, ${\bf r}=0$, the bound state contribution to 
the spin density $<s({\bf r}=0 )>$ changes at the transition 
point from the up spin to the down spin. So there exists the 
discontinuity of $<s({\bf r}=0 )>$ at the transition. 
But the main contribution of the $<s({\bf r}=0 )>$ comes from 
the continuum part of the spectrum.
\section{Conclusion and Discussion}
In this paper we have  investigated 
the non-magnetic impurity effect by 
solving the BdG equation numerically for chiral 
$p_{x}+{\rm i}p_{y}$ and the non-chiral but 
${\cal T}$-violating $d_{xy}+{\rm i}s$ superconductor in $\S 2$.
We see  induced currents of the $p_{x}+{\rm i}p_{y}$ and 
$d_{xy}+{\rm i}s$ superconductors. Both of the 
${\cal T}$-violating superconductor can generate the induced 
spontaneous currents 
around the impurities. The induced current can 
generate localized magnetic field 
in the superconductor. But the patterns of
the induced currents 
are quite different reflecting 
the property of the Cooper pairs. 
In the chiral superconductor $p_{x}+{\rm i}p_{y}$, 
circular currents around the impurity are 
induced spontaneously. On the other hand, in 
the $d_{xy}+{\rm i}s$ superconductor,  
the current pattern has quadrapolar shape   
reflecting the mixture of the angular 
momentum ${\rm L}_{z}=2$ and ${\rm L}_{z}=-2$. \\
In $\S 3$ we investigate the magnetic impurity in 
$p_{x}+{\rm i}p_{y}$ superconductor. In this case, the circular 
current is induced as in the non-magnetic impurity case. 
But the bound states also carry the current in contrast to 
the non-magnetic case.
The level of the bound state is not different as in the 
non-magnetic case and the phase transition of the ground 
state is absent when the `ideal' conditions, such as 
electron-hole symmetry and s-wave scattering from the impurity, 
are satisfied in the p-wave case. 
 But if some conditions are violated, then 
there is a transition to the electron spin-polarized 
state. Especially, we pay attention to the anisotropic scattering 
from the impurity which cause the phase transition to the 
spin polarized state. \\
In this paper, we take a classical impurity spin 
and neglect the transversal components 
of the magnetic impurity spin.
So we do not consider  the Kondo 
effect is the usual sense.
It is expected that in the ${\cal T}$ violating 
superconductor the induced current  pins the impurity 
spin to the parallel direction of the generated magnetic field. 
This effect would weaken the Kondo effects in addition to 
the presence of the superconducting gap. 
But if the Zeeman energy is small, then the 
Kondo effect should be taken into account.
It is interesting to 
investigate the competition of the Kondo effect 
and this Zeeman energy in ${\cal T}$-violating 
superconductor. Other than this effect, in the triplet 
superconductor, there are many interesting effects in 
connection to the Kondo effect. 
Especially due to the lack of 
the spin symmetry of the superconducting state and the 
mixture of the finite angular momentum channels 
other than s-state (angular momentum $m=0$ channel),
the Kondo effect in the triplet superconductor 
may have different features from the s-wave case. 
This problem is left for future studies.
Also the spin-orbit interaction shows  interesting 
effects. The spin-orbit interaction directly couples to the 
orbital part of the Cooper pair and give different effects 
for the $p_{x}+{\rm i}p_{y}$ and $p_{x}-{\rm i}p_{y}$ 
state. So the induced components, $p_{x}-{\rm i}p_{y}$, 
around the impurity in $p_{x}+{\rm i}p_{y}$ superconductor 
will be affected by the spin-orbit interaction. Also this 
problem will be discussed elsewhere.
\section*{Acknowledgment}
The author is grateful to  M. Sigrist for many helpful 
discussion and suggestions. He also thanks to M. Matsumoto 
and Y. Onishi  for  discussion and K. K. Ng for  
reading this manuscript
Numerical computation in this work was carried out at the 
Yukawa Institute Computer Facility. The author has been 
supported by COE fellowship from the Ministry of Education, 
Science, Sports and Culture of Japan. 
\section*{Appendix 1}
Electron-hole asymmetry shifts the 
the energy levels of the bound states. 
Here we concentrate the $p_{x}+{\rm i}p_{y}$ type of 
Cooper pairs. As for impurity potential, we take the 
contact type s-wave scattering like 
${\rm U}\delta ({\bf r})$.
The Fourier transformed BdG equation is written as,
\begin{eqnarray}
\xi_{k} u_{k \sigma}+ \Delta_{k}v_{k -\sigma } 
+\frac{{\rm U}}{V} \sum_{k}u_{k \sigma } &=& \epsilon u_{k \sigma} \\
-\xi_{k}v_{k -\sigma}+ \Delta^{\ast}_{k}u_{k \sigma } 
-\frac{{\rm U}}{V} \sum_{k}v_{k -\sigma } &=& \epsilon v_{k -\sigma} 
\label{Bdguni}
\end{eqnarray}
where $u_{k}$ and $v_{k}$ are electron and hole part of the 
wave function, respectively, 
$\xi_{k}$ is quasi-particle dispersion and $\Delta_{k}$ 
is gap function. We introduce the variable 
${\rm I}_{0} =({\rm U}/V) \sum_{k}u_{k \sigma }$ 
and ${\rm I}_{0}{}^{'} =({\rm U}/V) \sum_{k}v_{k -\sigma }$, 
and the self-consistent 
equation for ${\rm I}_{0}$ and ${\rm I}_{0}{}^{'}$ are given as
\begin{equation} \begin{array}{l} \displaystyle 
{\rm I}_{0} = \frac{{\rm U}}{V} \sum_{k}u_{k \sigma}=
\frac{{\rm U}}{V} \sum_{k} \frac{(\epsilon +\xi_{k}){\rm I}_{0}
-\Delta_{k} {\rm I}_{0}{}^{'}}
{\epsilon^{2}-\xi_{k}{}^{2}-|\Delta_{k}|{}^{2} }
= {\rm I}_{0} \frac{{\rm U}}{V} \sum_{k} 
\frac{\epsilon +\xi_{k}}
{\epsilon^{2}-\xi_{k}{}^{2}-|\Delta_{k}|{}^{2} }  \\ \\
\displaystyle 
{\rm I}_{0}{}^{'} = \frac{{\rm U}}{V} \sum_{k}v_{k -\sigma} = 
\frac{{\rm U}}{V}\sum_{k} \frac{\Delta^{\ast}{}_{k}{\rm I}_{0} 
-(\epsilon -\xi_{k}){\rm I}_{0}{}^{'}}
{\epsilon^{2}-\xi_{k}{}^{2}-|\Delta_{k}|{}^{2} } = 
{\rm I}_{0}{}^{'}\frac{{\rm U}}{V} \sum_{k} 
\frac{ -\epsilon+\xi_{k}}
{\epsilon^{2}-\xi_{k}{}^{2}-|\Delta_{k}|{}^{2} } \\ 
\label{bound1}
\end{array} \end{equation}
where for final form we used the fact that the angular 
integral over the gap function $\Delta_{k}$ vanishes. 
The $\xi_{k}$ term in the numerator does not vanish 
due to the electron-hole asymmetry.
The solutions of (\ref{bound1}) are divided into two types 
(${\rm I}_{0} \ne 0$ and ${\rm I}_{0}{}^{'}=0$ ) and 
(${\rm I}_{0}=0$ and ${\rm I}_{0}{}^{'} \ne 0$ ).
Here we introduce the electron-hole asymmetry 
phenomenologically like 
${\rm N}(\xi )={\rm N}_{0}+{\rm N}^{'}\xi $, 
where ${\rm N}(\xi )$ is the density of state.
Then with neglecting the $k$-dependence of the 
$|\Delta_{k}|{}^{2}$ in denominator of (\ref{bound1}), 
which are approximately valid in the $p_{x}+{\rm i}p_{y}$ case, 
in the unitary limit ${\rm U} \rightarrow \infty$, 
the energy level of the bound state (
$\epsilon < |\Delta_{k} |$ ) is given like,  
\begin{eqnarray}
\epsilon &\sim & \pm \frac{|2\Delta {\rm N}^{'}D|}
{\sqrt{({\rm N}_{0}\pi )^{2}+(2{\rm N}^{'}D)^{2}}} 
\end{eqnarray} 
where $D$ is the half of the band-width.
The energy level 
 does not coincide to zero energy level but splits 
due to the electron-hole asymmetry. The above argument is 
also applicable to the d-wave superconductor.
\section*{Appendix 2}
In this Appendix we discuss the 
magnetic field generated around the impurity in 
the $d+{\rm i}s$ type Cooper pair. Our discussion is 
based on the GL theory. 
In the case of the coexistence of the 
d and s wave with ${\cal T}$-violating superconductor, the GL 
free energy in the weak coupling approximation is written as 
\begin{eqnarray}
&&f = -2\alpha_{s} |\Delta_{s} |^{2} 
-\alpha_{d} |\Delta_{d} |^{2} + 
\beta (|\Delta_{s} |^{4} + \frac{3}{8}|\Delta_{d} |^{4} + 
2|\Delta_{s} |^{2}|\Delta_{d} |^{2} + 
\frac{1}{2}[\Delta_{s}{}^{*}{}^{2}\Delta_{d}{}^{2}
+ \Delta_{s}{}^{2}\Delta_{d}{}^{*}{}^{2})] \nonumber \\
&&+ K[ 2|{\bf \Pi} \Delta_{s}{}^{*}|+|{\bf \Pi} \Delta_{d}{}^{*}| +
( \Pi_{x}{}^{*}\Delta_{s}\Pi_{x}\Delta_{d}{}^{*}-
\Pi_{y}{}^{*}\Delta_{s}\Pi_{y}\Delta_{d}{}^{*}+
{\rm c.c})]
+g(|\Delta_{s} |^{2}+|\Delta_{d} |^{2})\delta (r) \nonumber \\
&&
\end{eqnarray}
where $\alpha_{s}={\rm ln}\frac{{\rm T}_{cs}}{{\rm T}} $ and 
$\alpha_{d}={\rm ln}\frac{{\rm T}_{cd}}{{\rm T}} $ 
and we set the same strength 
of the impurity potential for d and s wave order parameter.  \\
The derived GL equations are,
\begin{eqnarray}
-2\alpha S+\frac{8}{3}(|S|^{2}+|D|^{2})S
+\frac{3}{4}D^{2}S^{*}
+2\Pi^{*}{}^{2}S+
(\Pi^{2}{}_{x}-\Pi^{2}{}_{y})D+gS\delta (r) = 0 && \\
-D+\frac{4}{3}(\frac{3}{4}|D|^{2}+2|S|^{2})D+
\frac{4}{3}S^{2}D^{*}
+\Pi^{*}{}^{2}D+
(\Pi^{2}{}_{x}-\Pi^{2}{}_{y})S+gD\delta (r) = 0 && \\
j = [ 2S^{*}{\bf \Pi}S+D^{*}{\bf \Pi}D+
(S\Pi_{x}D^{*}+D\Pi_{x}S^{*})\hat{x}
-(S\Pi_{y}D^{*}+D\Pi_{y}S^{*})\hat{y}+{\rm c.c}] &&  
\end{eqnarray}
here $\alpha = \frac{\alpha_{s}}{\alpha_{d}}$ and 
${\bf \Pi} = -{\rm i}{\bf \nabla}$. $D$ and $S$ is 
the order parameters for d- ans s-wave, respectively, which 
are normalized by  
$\Delta_{0}=\sqrt{\frac{4\alpha_{d}}{3\beta}}$ and 
the length scale is also normalized 
by $\sqrt{\frac{K}{\alpha_{d}}}$. 
For the realization of the bulk ${\cal T}$-violating state 
we  impose the condition $1 > \alpha > \frac{2}{3}$.
We expand $S$ and $D$ in first 
order of impurity potential,  
$g$, like $S={\rm i}(S_{0}+S_{1})$ and 
$D=(D_{0}+D_{1})$, where $S_{0}$ and $D_{0}$ are the values 
of the homogeneous case and are given as 
$S_{0}{}^{2}=\frac{3}{4}(3\alpha -2)$ and 
$D_{0}{}^{2}=3(1-\alpha )$.
The Fourier transformed form of  $S_{1}$ and $D_{1}$ 
are rather complex and are given like
\begin{eqnarray}
S^{'}(q,\phi ) &=&
\frac{2\,{\sqrt{-6 + 9\,\alpha}}\,g\,
     \left( 16\,{\alpha^2} + 
     3\,{q^2}\,\left( -4 + {q^2} \right)  - 
      2\,\alpha \,\left( 8 - 9\,{q^2} \right)  - 
      {q^2}\,\left( 4 - 2\,\alpha + {q^2} \right) 
      \,\cos (4\,\phi) \right) }
{C(q,\phi)} \nonumber \\
S^{''}(q,\phi ) &=& 
\frac{4\,{\sqrt{3 - 3\,\alpha}}\,g\,\cos (2\,\phi)\,
     \left( 16 + 48\,{\alpha^2} - 16\,{q^2} 
     + 3\,{q^4} + 8\,\alpha \,\left( -7 + 3\,{q^2} \right)
     - {q^4}\,\cos (4\,\phi) \right) }
{C(q,\phi)} \nonumber \\
D^{'}(q,\phi ) &=& 
\frac{8{\sqrt{3 - 3\,\alpha}}\,g\,\left( 
-16\,\alpha + 24\,{\alpha^2} - 10\,{q^2} + 
       21\,\alpha \,{q^2} + 3\,{q^4} - 
       {q^2}\,\left( 2 - \alpha + {q^2} \right) 
       \,\cos (4\,\phi) \right) }
{C(q,\phi)} \nonumber \\
D^{''}(q,\phi ) &=& 
\frac{4{\sqrt{-6 + 9\,a}}\,g\,\cos (2\,\phi)\,
     \left( -2\,\left( 8 + 16\,{\alpha^2} 
     - 6\,{q^2} + {q^4} + 
     6\,\alpha \,\left( -4 + {q^2} \right)  \right)  + 
       {{q}^4}\,{{\cos (2\,\phi )}^2} \right) }
 {C(q,\phi)} \label{DSfirst} 
\end{eqnarray}
where we set $(q_{x}, q_{y})=(q\cos (\phi),q\sin (\phi) )$ 
and decompose like $S_{1}=S^{'}+{\rm i}S^{''}$ and 
$D_{1}=D^{'}+{\rm i}D^{''}$.
The denominator factor, $C(q,\phi )$, is given as, 
\begin{eqnarray}
C(q,\phi ) &=&
512\alpha - 
1280{\alpha^2} + 768{\alpha^3} + 
320{q^2} - 896\alpha {q^2} + 
480{\alpha^2}{q^2} - 48{q^4}  
-48\alpha {q^4} - 19{q^6}   \nonumber \\
&+& 4{q^2}\,\left( 16 + 24{\alpha^2} + 4{q^2} + 3{q^4} +
4\alpha \left( -8 + {q^2} \right)  \right) \cos (4\phi ) 
- {q^6}\cos (8\phi ) 
\end{eqnarray}
The current in the first order of $g$ is given as 
\begin{eqnarray}
{\bf j}({\bf r}) &=& (-4S_{0}\partial_{x}S^{''}({\bf r})
-2D_{0}\partial_{x}D^{''}({\bf r})
+2S_{0}\partial_{x}D^{'}({\bf r})-
2D_{0}\partial_{x}S^{'}({\bf r}))\hat{{\bf x}} \\
&+& 
(-4S_{0}\partial_{y}S^{''}({\bf r})
-2D_{0}\partial_{y}D^{''}({\bf r})
+2S_{0}\partial_{y}D^{'}({\bf r})-
2D_{0}\partial_{y}S^{'}({\bf r}))\hat{{\bf y}}
\end{eqnarray}
The magnetic field can be calculated by the Biot-Savart 
formula like,
\begin{eqnarray}
B(q) &=&  \frac{{\bf q}\times {\bf j}({\bf q})}{q^{2}} \nonumber \\
&=& -\frac{q_{x}q_{y}(S_{0}D^{'}(q)-D_{0}S^{'}(q))}{q^{2}} \label{magfid} 
\end{eqnarray}
We can easily see the sign of the magnetic field changes 
by 90 degree rotation in $(q_{x},q_{y})$ space 
from  (\ref{DSfirst}) and (\ref{magfid}), so the feature of the 
generated magnetic field is the quadrapole like. 
The feature is consistent to our numerical calculations in $\S$2. 
\section*{Appendix 3}
The energy levels of the 
bound states with p-wave scattering from the impurity 
are determined as below. We take the same form of the 
magnetic impurity potential as (\ref{magpoten}). 
Then as in the Appendix 1 from the Fourier transformed form of 
the BdG equation we get 
\begin{subeqnarray}
u_{\uparrow,N}(k) &=& \frac{(\epsilon_{N}+\xi_{k})
({\rm I}_{0}+{\rm I}_{1}{\rm e}^{{\rm i}\phi}
+{\rm I}^{'}{}_{1}{\rm e}^{-{\rm i}\phi})
+\Delta_{k}({\rm J}_{0}+{\rm J}_{1}{\rm e}^{{\rm i}\phi}
+{\rm J}{}^{'}{}_{1}{\rm e}^{-{\rm i}\phi})}
{\epsilon^{2}{}_{N}-\xi^{2}{}_{k}-\Delta_{0}{}^{2}} \\
v_{\downarrow,N}(k) &=& 
\frac{\Delta^{\ast}{}_{k}({\rm I}_{0}
+{\rm I}_{1}{\rm e}^{{\rm i}\phi}
+{\rm I}^{'}{}_{1}{\rm e}^{-{\rm i}\phi})+
(\epsilon_{N}-\xi_{k})({\rm J}_{0}+
{\rm J}_{1}{\rm e}^{{\rm i}\phi}
+{\rm J}^{'}{}_{1}{\rm e}^{-{\rm i}\phi})}
{\epsilon^{2}{}_{N}-\xi^{2}{}_{k}-\Delta_{0}{}^{2}}, \label{uvexp}
\end{subeqnarray}
here we define as 
${\rm I}_{0}={\rm J_{s}S}\sum_{k^{'}}u_{\uparrow,N}(k^{'})$, 
${\rm I}_{1}={\rm J_{p}S}\sum_{k^{'}}u_{\uparrow,N}(k^{'})
{\rm e}^{-{\rm i}\phi^{'}}$,
${\rm I}{}^{'}_{1}={\rm J_{p}S}\sum_{k^{'}}u_{\uparrow,N}(k^{'})
{\rm e}^{{\rm i}\phi^{'}}$,  
${\rm J}_{0}={\rm J_{s}S}\sum_{k^{'}}v_{\downarrow,N}(k^{'})$, 
${\rm J}_{1}={\rm J_{p}S}\sum_{k^{'}}v_{\downarrow,N}(k^{'})
{\rm e}^{-{\rm i}\phi^{'}}$ and 
${\rm J}{}^{'}_{1}={\rm J_{p}S}\sum_{k^{'}}v_{\downarrow,N}(k^{'})
{\rm e}^{{\rm i}\phi^{'}}$, where $\phi^{'}$ is the angle 
of $k^{'}$. 
Assuming $|\epsilon_{N} |< \Delta_{0}$, 
we get the bound energies 
from the linear equations 
of ${\rm I}_{0}$,${\rm I}_{1}$,${\rm I}^{'}{}_{1}$,${\rm J}_{0}$,
${\rm J}_{1}$ and ${\rm J}^{'}{}_{1}$. The energy levels of the 
bound states with the impurity potential (\ref{magpoten2}) are given 
by replacing ${\rm J}_{0}$ to $-{\rm J}_{0}$ in the above 
expression (\ref{uvexp}).
 
\end{document}